\documentclass[twocolumn]{aastex63}

\usepackage[T1]{fontenc}

\DeclareRobustCommand{\VAN}[3]{#2}
\let\VANthebibliography\thebibliography
\def\thebibliography{\DeclareRobustCommand{\VAN}[3]{##3}\VANthebibliography}

\usepackage{graphicx}	
\usepackage{amsmath}	
\usepackage{amssymb}	

\usepackage{hyperref}
\usepackage{academicons}
\usepackage{xcolor}

\usepackage{tablefootnote}

\usepackage{lineno}

\usepackage{newtxtext,newtxmath}

\newcommand{\orcid}[1]{\href{https://orcid.org/#1}
    {\includegraphics[scale=0.08]{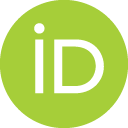}}
}
\newcommand{\tess}{\emph{TESS}}
\newcommand{\juliet}{\texttt{Juliet}}
\newcommand{\batman}{\texttt{batman}}
\newcommand{\esinw}{\sqrt{e}\sin\omega}
\newcommand{\ecosw}{\sqrt{e}\cos\omega}

\graphicspath{{./}{Figures/}}

\date{Accepted XXX. Received YYY; in original form ZZZ}

\begin{document}

\title[Validation of TOI-1221 b]{Validation of TOI-1221 b: A warm sub-Neptune exhibiting TTVs around a Sun-like star}

\shorttitle{Validation of TOI-1221 b} 
\shortauthors{Mann et al.}

\correspondingauthor{Christopher Mann}
\email{christopher.mann@umontreal.ca}



\author[0000-0002-9312-0073]{Christopher Mann}
\affiliation{Université de Montréal}
\affiliation{Trottier Institute for Research on Exoplanets (\emph{iREx})}

\author[0000-0002-6780-4252]{David Lafrenière}
\affiliation{Université de Montréal}
\affiliation{Trottier Institute for Research on Exoplanets (\emph{iREx})}

\author[0000-0003-2313-467X]{Diana Dragomir}
\affiliation{Department of Physics and Astronomy, University of New Mexico, 1919 Lomas Blvd NE Albuquerque, NM, 87131, USA}

\author[0000-0002-8964-8377]{Samuel N. Quinn} 
\affiliation{Center for Astrophysics \textbar \ Harvard \& Smithsonian, 60 Garden Street, Cambridge, MA 02138, USA}

\author[0000-0001-5603-6895]{Thiam-Guan Tan} 
\affiliation{Perth Exoplanet Survey Telescope, Perth, Western Australia, Australia}

\author[0000-0001-6588-9574]{Karen A.\ Collins} 
\affiliation{Center for Astrophysics \textbar \ Harvard \& Smithsonian, 60 Garden Street, Cambridge, MA 02138, USA}

\author[0000-0002-2532-2853]{Steve~B.~Howell}
\affil{NASA Ames Research Center, Moffett Field, CA 94035, USA}

\author{Carl Ziegler} 
\affiliation{Department of Physics, Engineering and Astronomy, Stephen F. Austin State University, 1936 North St, Nacogdoches, TX 75962, USA}

\author{Andrew W. Mann} 
\affiliation{Department of Physics and Astronomy, The University of North Carolina at Chapel Hill, Chapel Hill, NC 27599-3255, USA}

\author[0000-0002-3481-9052]{Keivan G.\ Stassun}
\affiliation{Department of Physics and Astronomy, Vanderbilt University, Nashville, TN 37235, USA} 

\author[0000-0002-2607-138X]{Martti~H.~Kristiansen}
\affil{Brorfelde Observatory, Observator Gyldenkernes Vej 7, DK-4340 T\o{}ll\o{}se, Denmark}

\author{Hugh Osborn}
\affiliation{Department of Physics and Kavli Institute for Astrophysics and Space Research, Massachusetts Institute of Technology, Cambridge, MA 02139, USA}
\affiliation{Physikalisches Institut, University of Bern, Gesellsschaftstrasse 6, 3012 Bern, Switzerland}

\author[0000-0001-9879-9313]{Tabetha Boyajian} 
\affiliation{Louisiana State University 202 Nicholson Hall Baton Rouge, LA 70803, USA}

\author{Nora Eisner}
\affiliation{Sub-department of Astrophysics, University of Oxford, Keble Rd, Oxford, United Kingdom}

\author{Coel Hellier}
\affiliation{Astrophysics Group, Keele University, Staffordshire ST5 5BG, U.K.}

\author[0000-0003-2058-6662]{George~R.~Ricker} 
\affiliation{Department of Physics and Kavli Institute for Astrophysics and Space Research, Massachusetts Institute of Technology, Cambridge, MA 02139, USA}

\author[0000-0001-6763-6562]{Roland~Vanderspek} 
\affiliation{Department of Physics and Kavli Institute for Astrophysics and Space Research, Massachusetts Institute of Technology, Cambridge, MA 02139, USA}

\author[0000-0001-9911-7388]{David~W.~Latham} 
\affiliation{Center for Astrophysics \textbar \ Harvard \& Smithsonian, 60 Garden Street, Cambridge, MA 02138, USA}

\author[0000-0002-6892-6948]{S.~Seager} 
\affiliation{Department of Physics and Kavli Institute for Astrophysics and Space Research, Massachusetts Institute of Technology, Cambridge, MA 02139, USA}
\affiliation{Department of Earth, Atmospheric and Planetary Sciences, Massachusetts Institute of Technology, Cambridge, MA 02139, USA}
\affiliation{Department of Aeronautics and Astronautics, Massachusetts Institute of Technology, Cambridge, MA 02139, USA}

\author[0000-0002-4265-047X]{Joshua~N.~Winn} 
\affiliation{Department of Astrophysical Sciences, Princeton University, 4 Ivy Lane, Princeton, NJ 08544, USA}

\author[0000-0002-4715-9460]{Jon~M.~Jenkins} 
\affiliation{NASA Ames Research Center, Moffett Field, CA}

\author[0000-0002-4625-8264]{Jesus Noel Villaseñor}
\affiliation{Department of Physics and Kavli Institute for Astrophysics and Space Research, Massachusetts Institute of Technology, Cambridge, MA 02139, USA}

\author[0000-0002-8058-643X]{Brian~McLean} 
\affiliation{Space Telescope Science Institute, 3700 San Martin Drive, Baltimore, MD, 21218, USA}

\author[0000-0002-4829-7101]{Pamela~Rowden} 
\affiliation{Royal Astronomical Society, Burlington House, Piccadilly, London W1J 0BQ, UK}

\author[0000-0002-5286-0251]{Guillermo~Torres} 
\affiliation{Center for Astrophysics \textbar \ Harvard \& Smithsonian, 60 Garden Street, Cambridge, MA 02138, USA}

\author[0000-0003-1963-9616]{Douglas A. Caldwell} 
\affiliation{NASA Ames Research Center, Moffett Field, CA}
\affiliation{SETI Institute}

\author[0000-0003-2781-3207]{Kevin I.\ Collins} 
\affiliation{George Mason University, 4400 University Drive, Fairfax, VA, 22030 USA}

\author[0000-0001-8227-1020]{Richard P. Schwarz} 
\affiliation{Center for Astrophysics \textbar \ Harvard \& Smithsonian, 60 Garden Street, Cambridge, MA 02138, USA}

\begin{abstract}
	We present a validation of the long-period ($91.68278^{+0.00032}_{-0.00041}$ days) transiting sub-Neptune planet \object[TIC349095149.01]{TOI-1221~b} (TIC 349095149.01) around a Sun-like (m$_{\rm V}$=10.5) star.   
	This is one of the few known exoplanets with period >50 days, and belongs to the even smaller subset of which have bright enough hosts for detailed spectroscopic follow-up.
	We combine {\it TESS} light curves and ground-based time-series photometry from PEST (0.3~m) and LCOGT (1.0~m) to analyze the transit signals and rule out nearby stars as potential false positive sources.
	High-contrast imaging from SOAR and Gemini/Zorro rule out nearby stellar contaminants.
	%
	Reconnaissance spectroscopy from CHIRON sets a planetary scale upper mass limit on the transiting object (1.1 and 3.5 M$_{\rm Jup}$ at 1$\sigma$ and 3$\sigma$, respectively) and shows no sign of a spectroscopic binary companion. 
	We determine a planetary radius of
    $R_{\rm p} = 2.91^{+0.13}_{-0.12} R_{\earth}$, placing it in the sub-Neptune regime.
	%
	With a stellar insolation of 
    $S = 6.06^{+0.85}_{-0.77}\ S_{\earth}$, 
	we calculate a moderate equilibrium temperature of 
    $T_{\rm eq} =$ 440 K, 
    assuming no albedo and perfect heat redistribution.
    We find a false positive probability from TRICERATOPS of FPP $ = 0.0014 \pm 0.0003$ as well as other qualitative and quantitative evidence to support the statistical validation of TOI-1221 b.
    We find significant evidence 
    (>$5\sigma$)
    of oscillatory transit timing variations, likely indicative of an additional non-transiting planet.
    %
\end{abstract}




\section{Introduction}

In July of 2018, the Transiting Exoplanet Survey Satellite (\tess{}) mission \citep{TESS_2015} began its task of observing nearly the entire sky for exoplanets transiting around bright and nearby stars.  
Although the {\it Kepler} mission \citep{Kepler_2010} revolutionized exoplanet science with its explosion of discoveries, most of the {\it Kepler} targets are quite faint.  
The brightness of {\it TESS} host stars allows for much more detailed follow up in terms of mass measurement and atmospheric characterization.  At the time of writing, the mission has produced
6133 {\it TESS} Objects of interest (TOIs), of which 1646 have been identified as false positives and only 282
confirmed as \emph{bona fide} exoplanets \citep{TESS_website}.  
The follow-up work required to verify and confirm the still-growing list of \tess{} candidates is substantial, challenging, and yet absolutely essential to the overall success of the mission.

Not all exoplanet systems are equal in their ease of detection and characterization.  Those with deeper transits (i.e. large planets and/or small host stars) produce strong signals in their light curves and shorter-period planets have both higher transit probabilities and more frequent transit events that require less baseline time coverage to observe.  
In direct contrast, long-period planets are both harder to initially detect, and allow for only infrequent follow-up opportunities.  The vast majority of known exoplanets for which we have both reliable mass and radius measurements have periods $\lesssim 50$ days \citep{NEXA_Planetary_Systems_Table}.

\begin{figure*}
	\centering
	\includegraphics[width=0.99\textwidth]
    {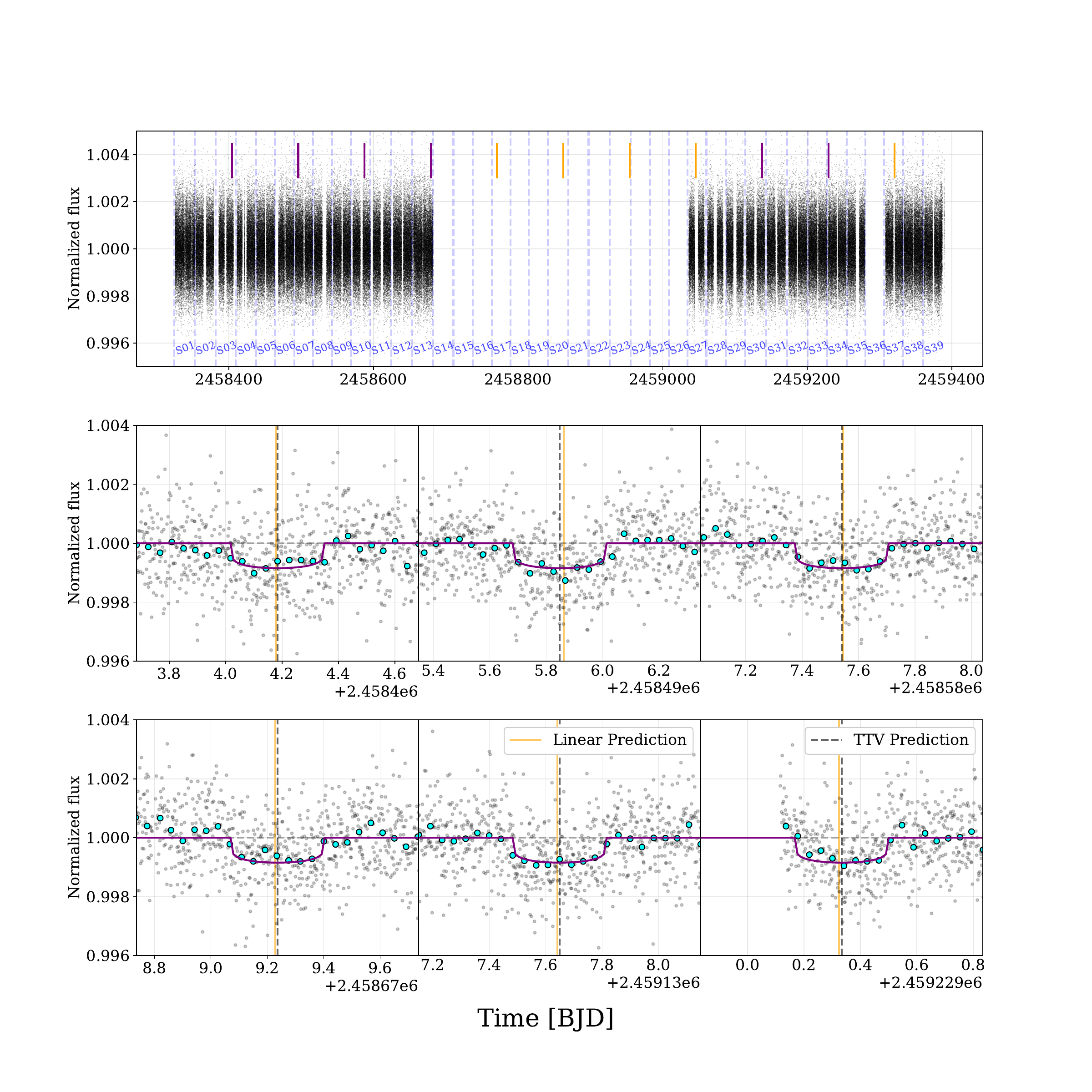}
    \caption{
    \emph{First row:} Sector-by-sector PDCSAP flux, normalized to within-sector median.  Purple bars indicate where transits were detected, orange bars indicate where expected transits fell in observing gaps. 
    \emph{Rows 2 \& 3:} Zoomed plot of each of the six caught transits.  The {\it TESS} data were taken at a cadence of 2 minutes and are shown binned to 60-minute intervals with the cyan points.  The displayed transit model is described in Sections~\ref{Sec:analysis} and \ref{Sec:discussion}.  
    Orange vertical lines show the predicted midpoints given a fixed period ephemeris.  Dashed black lines indicate the fitted midpoints, showing TTVs described in Section~\ref{sSec:Transit_Fitting}.
    \label{fig:TESS_data}}
\end{figure*}


Despite the poorer statistics on smaller, cooler exoplanets there has arisen a clear bimodal distribution of radii split on either side of $\sim$2~Earth radii that also has a dependence on the planet’s insolation flux \citep{Fulton_2017}.  There are competing ideas as to why the super-Earth/sub-Neptune valley occurs (e.g. photo-evaporation: 
\citet{Owen_Wu_2017,Jin_Mordasini_2018,Rogers_Owen_2021};
cooling-luminosity driven loss: 
\citet{Gupta_Schlichting_2019};
and gas-poor formation: 
\citet{Lee_Chiang_2016,Lopez_Rice_2018}).
There is still a major question of whether cool sub-Neptunes larger than 2~R$_{\earth}$ are mostly rocky or have sizeable H/He atmospheres.  
Stellar insolation's effect on atmospheric inflation may play a prominent role in distinguishing these options.
Vetting long-period exoplanet candidates provides new temperate exoplanet targets across a range of stellar insolation to help disentangle this confounding factor. 

In this paper we present the validation of \object[TIC349095149.01]{TOI-1221 b} (TIC 349095149.01), a sub-Neptune sized exoplanet located in the southern sky 
(RA: $07^{\rm h}11^{\rm m}41^{\rm s}.05$,
Dec: $-65^{\circ}30{\arcmin}31\farcs88$) around a Sun-like m$_{\rm V}$=10.5 star.
Having a wide 91.7 day orbit, this target is at the far edge of the known transiting long-period planet parameter space and its
atmospheric properties are less likely to be strongly driven by stellar insolation than its numerous hot counterparts.


The shallow transit signal in this system was initially noticed in the {\it TESS} Sector 7 data by the Visual Survey Group \citep[VSG;][]{Kristiansen_etal_2022} shortly after release.  Immediately checking back through previous sectors, a likely second transit was discovered in Sector 3, providing a tentative 91-day period.  
The candidate was also independently detected by the {\it TESS} Science Processing Operations Center \citep[SPOC;][]{SPOC_2016} at NASA Ames Research Center when conducting a multi-sector analysis of Sectors 1-9.  
The periodicity was confirmed shortly thereafter with a third transit detection in Sector 10 by the VSG.  
A later multi-sector SPOC analysis of Sectors 1-13 discovered a fourth transit and promoted the candidate to TOI status following a positive data validation report.

Since its discovery, the {\it TESS} Follow-up Observing Program (TFOP) community has made steady efforts to contribute key observations to confirm the planet's candidacy and rule out false positive scenarios.  
This manuscript aims to bring together and summarize the various contributed observations in order to validate this planet's legitimacy.
In Section~\ref{Sec:Observations} 
we provide an overview of the observations and data that contributed to the effort. 
Section~\ref{Sec:analysis} 
details the target's discovery, 
an assessment of false positive scenarios, and
stellar characterization.  It also describes the discovery of transit timing variations (TTVs) and the fitting procedures used to measure the system's parameters.  
Further discussion of the discovered planet properties and the implications of the TTVs  occurs in Section~\ref{Sec:discussion}. 
Finally, in Section~\ref{Sec:summary} 
we summarize the results of the entire planet validation process.

\section{Observations}\label{Sec:Observations}
In the effort to validate the target TOI-1221.01, data and observations were gathered from several teams and facilities and are summarized here.

\subsection{\tess{} light curves}

Due to its proximity to the southern ecliptic pole, TOI-1221 was observed by {\it TESS} over many sectors and with a very long baseline (more than 650 days of on-target coverage over 1065 days, at the time of writing).  It appeared in 25 Sectors (1-13, 27-35, and 37-39).  To date, six of those Sectors (3, 7, 10, 13, 30, and 34) caught a transit event allowing for a good transit fit and determination of the orbital parameters. It is scheduled to appear in 9 more sectors (61-69) in {\it TESS}'s Year 5 observations (early-to-mid 2023), and sectors 64 and 67 are expected to show additional transits given the ephemeris found later in this paper.  

We downloaded the Presearch Data Conditioning Simple Aperture Photometry 
\citep[PDCSAP;][]{Stumpe_etal_2012,Stumpe_etal_2014,Smith_etal_2012}
light curves from the Mikulski Archive for Space Telescopes (MAST) repository 
(doi:\dataset[10.17909/t9-nmc8-f686]{http://dx.doi.org/10.17909/t9-nmc8-f686}).
These data are background subtracted and have had instrumental systematics identified and removed.  The {\it TESS} data are visualized in Figure~\ref{fig:TESS_data}.

\begin{figure}%
	\centering
	\includegraphics[width=0.47\textwidth]
    {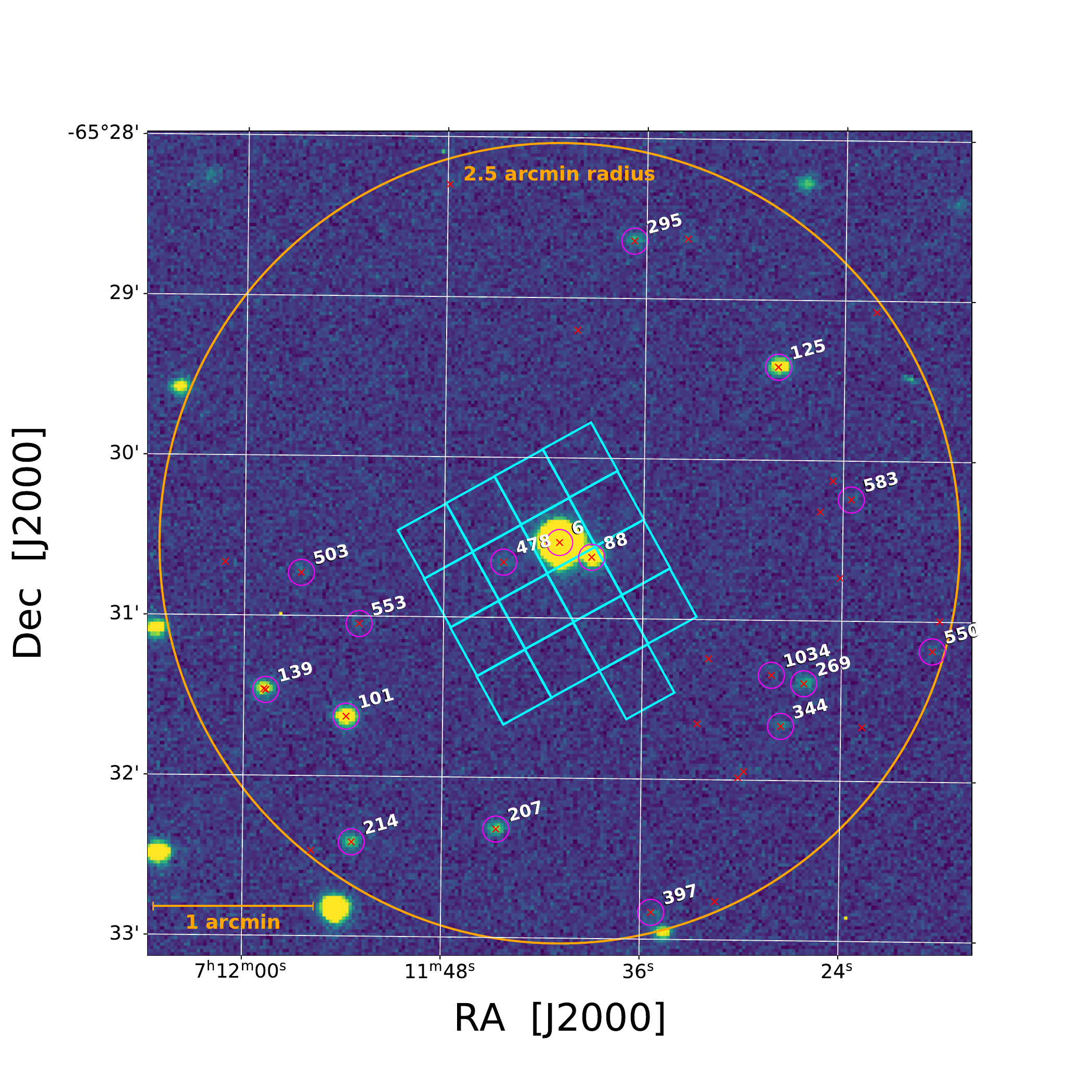}
    \caption{A 5\arcmin \ field of view from PEST centred on the target (centre object, ID \#6).  An example {\it TESS} aperture is overlaid in the same scale (cyan).  Identifier numbers come from a brightness-ordered list of {\it Gaia} stars (red X points) in the area.  Given {\it TESS}'s pixel size and PSF width, it is typical to check for NEBs within 2.5\arcmin of the target (orange circle). NEB check results for detectable neighbouring stars are displayed in Table~\ref{Tab:PEST_NEBs}.
        \label{fig:FOV}}
\end{figure}

\subsection{Ground-based light curves}

Follow-up observations were made by members of the TESS Follow-up Observing Program (TFOP) and are hosted on the Exoplanet Follow-up Observing Program (ExoFOP) TESS website 
(doi:\dataset[10.26134/ExoFOP3]{http://dx.doi.org/10.26134/ExoFOP3}).

\subsubsection{PEST}
We observed TOI-1221.01 on UTC 2021-01-14 in $\rm R_c$ band from the Perth Exoplanet Survey Telescope (PEST) near Perth, Australia. The $\sim$8.1 hour observation covered pre-ingress baseline through about 65\% of the predicted transit duration.  The 0.3 m telescope is equipped with a $1530\times1020$ SBIG ST-8XME camera with an image scale of 1$\farcs$2 pixel$^{-1}$ resulting in a $31\arcmin\times21\arcmin$ field of view. A custom pipeline based on {\tt C-Munipack}\footnote{http://c-munipack.sourceforge.net} was used to calibrate the images and extract the differential photometry, using an aperture with radius $5\farcs0$. The images have typical stellar point spread functions (PSFs) with a FWHM of $\sim 5\arcsec$.  The target star was intentionally saturated to check for fainter nearby eclipsing binaries (NEBs) near TOI-1221. The data rule out NEBs in stars within $2\farcm5$ of the target star that are fainter by as much as 7.5 magnitudes in $\rm R_c$ band, except for the nearest neighbor (TIC 349095148) at $13\farcs2$ separation which had a contaminated aperture.
An example PEST image is showin in Figure~\ref{fig:FOV}.

\begin{figure}
	\centering
    \includegraphics[width=0.47\textwidth]
    {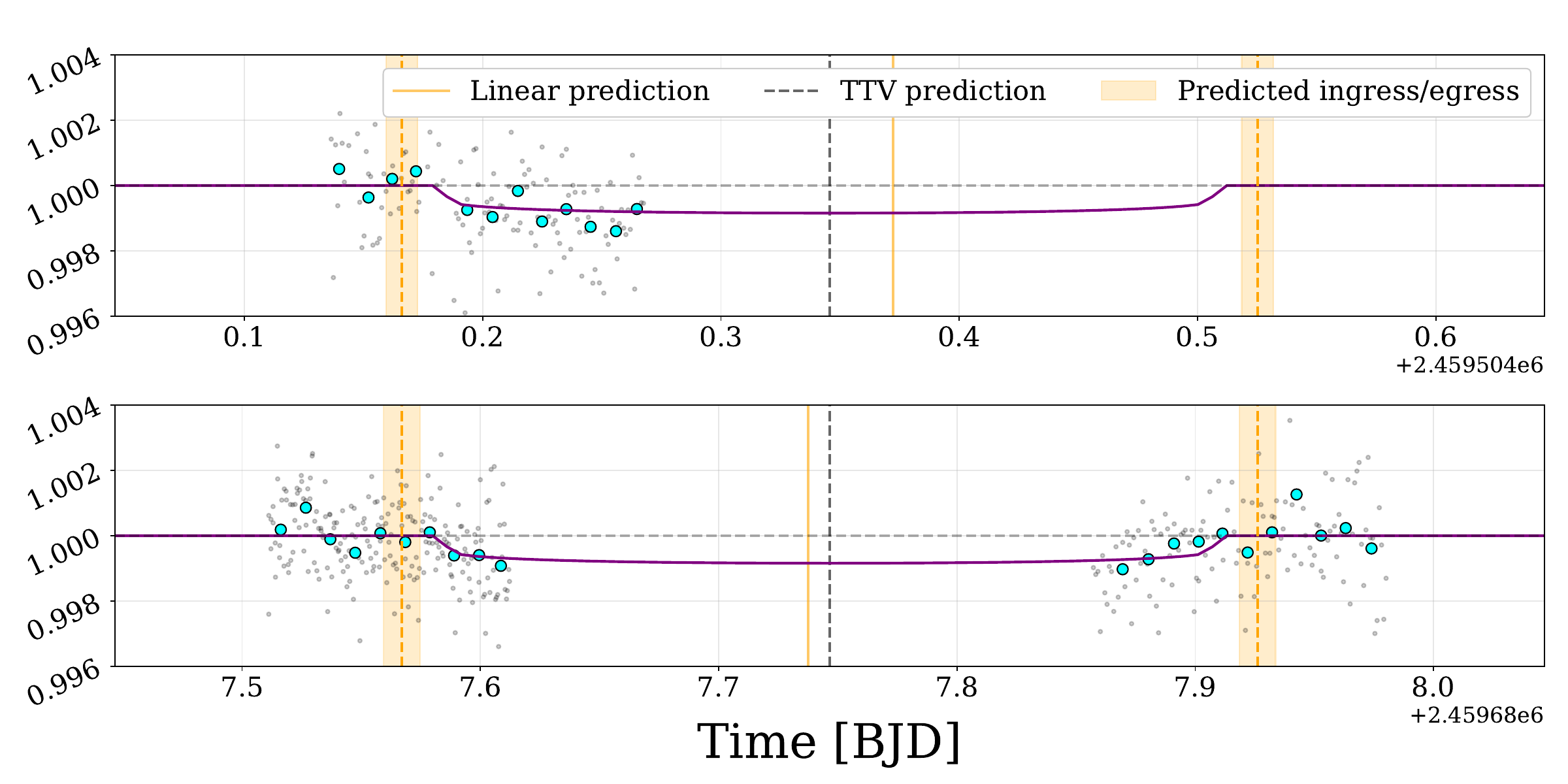}
    \caption{
    Light curves from the two LCOGT observations.  Cyan points are binned to 15-minute intervals.
    The predicted timings come from the fixed-period ephemeris of previous {\it TESS} data.
    The transit duration is long enough to require observing from two separate longitudes to detect both ingress and egress.
    Like in Figure~\ref{fig:TESS_data}, the central orange and black lines indicate the midpoint timing expected by a fixed-period model or fitted by an agnostic model allowing TTVs, respectively.
    }
     \label{fig:LCO_lc}
\end{figure}

\subsubsection{LCOGT}

In order to gather additional transit detections and to clear the NEB potential of the remaining 13.2\arcsec neighbour, we obtained two observations from the Las Cumbres Observatory global telescope network (LCOGT; \citet{LCO_paper}) 1.0~m telescopes using the SINISTRO instrument in the zs-band.  Being distributed in various locations around the globe, LCOGT observations of TOI-1221 were possible at a few southern sites: the Siding Springs Observatory (SSO) in Australia, the South African Astronomical Observatory (SAAO), and the Cerro Tololo Inter-American Observatory (CTIO) in Chile.
We used the {\tt TESS Transit Finder} tool, which is a customized version of the {\tt Tapir} software package \citep{Jensen_2013}, to schedule these LCOGT observations.

The first observation occurred on 
2021-10-16
where an ingress event was visible from the SSO site.  Simultaneous observations by two on-site 1.0~m telescopes ($A$ and $B$) were made, although substantial systematics on telescope $A$ prompted us to exclude it from further analysis.  The light curve from telescope $B$ achieved photometric precision comparable to the transit depth, and so was retained.  Unfortunately an attempted observation of the associated egress event at the SAAO site several hours later was thwarted by poor weather.

The second observation took place on 
2022-04-17.
This time the weather cooperated and we were able to capture both the ingress and egress events from the CTIO and SSO sites, respectively.  
The images from each observation were calibrated by the standard LCOGT {\tt BANZAI} pipeline \citep{McCully_etal_2018}. %
Photometric data were extracted using {\tt AstroImageJ} \citep{Collins_2017} and circular photometric apertures with radii $\sim5\farcs8$.  This aperture excludes most of the flux from the nearest known {\it Gaia} DR3 and TIC version 8 neighbor (TIC 349095148) which is  $13\farcs2$ SSW of TOI-1221. An aperture on TIC 349095148 also ruled out an NEB in this neighboring star (see section \ref{sec:RejectingFPs} and Table~\ref{Tab:PEST_NEBs}). 
%
LCOGT light curves are shown in Figure~\ref{fig:LCO_lc}.  

\begin{figure}
	\centering
	\includegraphics[width=0.47\textwidth]
    {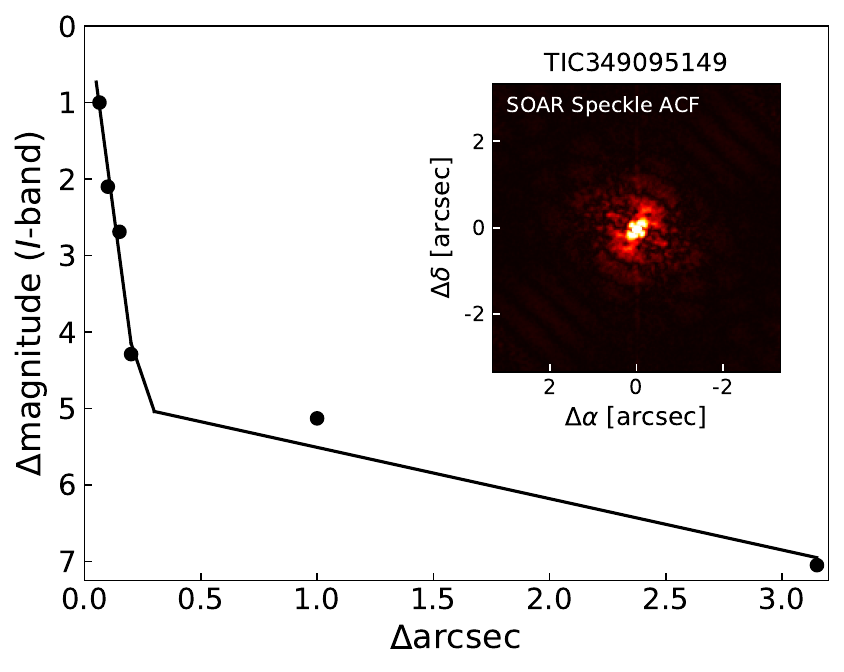}
	\includegraphics[width=0.47\textwidth]
    {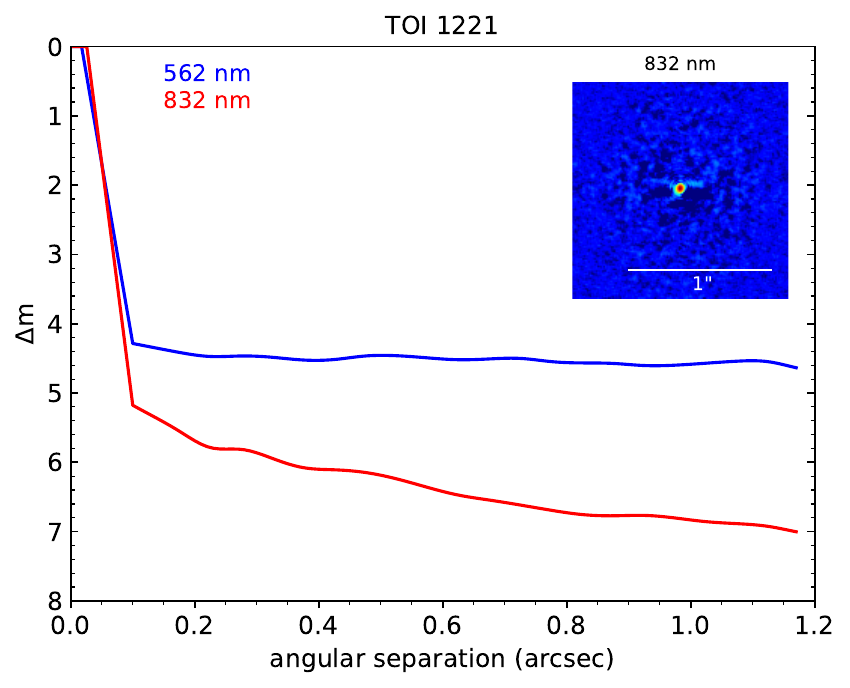}
    
    \caption{Contrast achieved via speckle imaging from SOAR (\emph{top}) and Gemini-S/Zorro (\emph{bottom}) for TOI-1221.  Neither measurement detects any neighbouring sources within the limits of their sensitivities ($5\sigma$).
        \label{fig:Speckle}}
\end{figure}

\subsection{WASP-South}



Though the transit depth of TOI-1221.01 is too shallow to be detectable by this instrument, the Wide Angle Search for Planets telescope (WASP; \citet{WASP_paper}) has archival data on this target that are useful to assess stellar activity and rotational modulation.  The archive contains 23,000 photometric data points on TOI-1221 spanning 4 observing seasons (2008/09, 2009/10, 2010/11 and 2011/12).
Typical cadence was 15 mins with individual exposures of 30 sec and a precision of about 6 mmag.



\begin{table}
\caption{NEB clearance by PEST and LCOGT}
\label{Tab:PEST_NEBs}
\begin{tabular}{ccccccc}
\hline
ID & $\Delta$mag &  RMS  & $\delta_{\rm NEB}$ & SNR & Target & Instr.\\
   &             & (ppt) & (ppt)              &     & sep.   & \\
\hline
101 & 4.03 &  17 &  20 & 14.6 & 1\arcmin43\arcsec & PEST \\
125 & 4.46 &  25 &  29 & 14.4 & 1\arcmin45\arcsec & PEST \\
139 & 4.82 &  30 &  41 & 17.0 & 2\arcmin03\arcsec & PEST \\
207 & 5.42 &  56 &  70 & 15.5 & 1\arcmin50\arcsec & PEST \\
214 & 5.43 &  58 &  71 & 15.2 & 2\arcmin16\arcsec & PEST \\
269 & 5.70 &  69 &  91 & 16.4 & 1\arcmin45\arcsec & PEST \\
295 & 5.77 &  76 &  98 & 16.0 & 1\arcmin56\arcsec & PEST \\
344 & 6.27 & 103 & 154 & 18.6 & 1\arcmin47\arcsec & PEST \\
397 & 6.34 & 189 & 165 & 10.8 & 2\arcmin23\arcsec & PEST \\
478 & 6.47 & 174 & 185 & 13.2 & 0\arcmin22\arcsec & PEST \\
503 & 7.36 & 331 & 422 & 15.8 & 1\arcmin37\arcsec & PEST \\
550 & 7.07 & 233 & 324 & 17.3 & 2\arcmin24\arcsec & PEST \\
553 & 7.16 & 219 & 350 & 19.8 & 1\arcmin22\arcsec & PEST \\
583 & 7.35 & 399 & 417 & 13.0 & 1\arcmin50\arcsec & PEST \\
1034& 7.30 & 418 & 399 & 11.8 & 1\arcmin32\arcsec & PEST \\
\hline
88  & 3.54 &  8.7 &  13 & 11.9 & 0\arcmin13\arcsec & LCOGT \\
\hline
\end{tabular}
\newline\noindent \textbf{Note.} The $\delta_{\rm NEB}$ quantity is a measure of how deep a transit would need to be on this particular target alone in order to cause the transit depth observed in the blended TESS aperture.  
Light curves for all these targets showed no indication of transit-like signals.  We use the simple transit SNR calculation of \citet[][eq.3]{Pont_etal_2006} to justify the non-detections on each neighbour star.  The aperture of the closest neighbour star (ID: 88) was contaminated by TOI-1221 in the PEST image and produced unreliable results.  The subsequent higher resolution LCOGT observation was able to resolve the two stars and perform a proper NEB check on this source.
\end{table}

\subsection{High-contrast imaging}
\subsubsection{SOAR}
Initial high-contrast speckle imaging was taken on 
2019-11-09
with the visiting HRCam on the Southern Astrophysical Research (SOAR) Telescope.  The HRCam is a fast imager that takes rapid diffraction-limited images with a Cousins \emph{I} filter and computes their power spectra and auto-correlation functions (ACFs).  Fringes in the power spectra or symmetric peaks in the ACFs indicate the detection of a nearby neighbour \citep{tokovinin_2018}.  The SOAR observation is shown in Figure~\ref{fig:Speckle} (top panel).

\subsubsection{Gemini-S/Zorro}
We acquired further high-contrast speckle imaging from Gemini South with the Zorro instrument on 
2021-10-19
(program ID: GN/S-2021A-LP-105) as part of the ongoing exoplanet follow-up imaging campaign of \citet{Howell_etal_2021}.  Zorro is able to simultaneously image in two filters, providing both 562 nm and an 832 nm measurements.  Seeing conditions were variable during the observation but remained within acceptable ranges.  Images were processed using the pipeline described in \citet{Howell_etal_2011}.
The Gemini/Zorro observation is also shown in Figure~\ref{fig:Speckle} (bottom panel).

\subsection{Reconnaissance Spectroscopy}
\subsubsection{CHIRON}

We obtained three spectra of TOI-1221 between 
2019-10-19
and 
2021-10-04
(program IDs: BOYA-19B-0232 and QUIN-21A-3268)
using the CHIRON high-resolution echelle spectrograph on the 1.5\,m SMARTS telescope at the CTIO \citep{Tokovinin2013}. In slicer mode, CHIRON uses a fiber-fed image slicer to feed the instrument and provides a resolving power of $R\sim80,000$\ across a broad wavelength range ($4100$--$8700$\,\AA). 
One spectrum was taken near transit, while the other two spectra were taken near opposite quadratures, providing the most radial velocity (RV) leverage.
These measurements are listed in Table~\ref{Tab:chiron_RVs}.
The CHIRON spectra can also be used to characterize the stellar properties, place conservative mass limits on the transiting planet, search for massive outer companions, and rule out false positive scenarios that would induce large RV variations or exhibit spectral line profile variations. 
We use the spectra provided by the CHIRON instrument team, optimally extracted according to the procedure described in \citet{Paredes2021}, which also demonstrates instrument velocity performance at the 
5 m s$^{-1}$
level for the brightest, slowly rotating, K dwarfs.


\section{Analysis}\label{Sec:analysis}

\subsection{Discovery}
The initial Sector 7 discovery of the TOI-1221.01 transit signal by the VSG \citep{Kristiansen_etal_2022} was made using \texttt{LcTools} \citep{Kipping_2015,Schmitt_2019} shortly after the data became available in early 2019.  The search was then extended to light curves of Sectors 1-7 from the Mikulski Archive for Space Telescopes (MAST), binned at 6 points per hour in order to search for additional transits. 
The second transit was located in Sector 3 giving a period estimate of 91.6 days. Once the Sector 10 data were released, a third matching transit was found, confirming the periodicity.

The transiting planet signature was also independently detected by the SPOC using an adaptive wavelet-based matched filter 
\citep{Jenkins_2002,Jenkins_etal_2010,Jenkins_etal_2020}
during a multi-sector search of sectors 1-9. The transit signature passed all the diagnostic tests performed by the Data Validation (DV) module \citep{Twicken_etal_2018} which also performed an initial limb-darkened model fit \citep{Li_etal_2019}. 
Further, the difference image centroiding analysis performed by the DV on sectors 1-39 localized the source of the transit signature to within $7\farcs6 \pm 6\farcs8$ of TOI-1221. The {\it TESS} Science Office (TSO) promoted this candidate to {\it TESS} Object of Interest status on 
2019-08-26 upon review of the DV report and other diagnostic information.
No signals from additional transiting planets were discovered in this target's light curve during these searches, nor from a box least squares search including the most recent sectors.

\subsection{False Positive Scenarios} \label{sec:RejectingFPs}

The most likely false positives to mimic an exoplanet in \tess{} observations involve transiting objects that are non-planetary in nature.  They may be physically bound to the host, or simply situated in close sky-projected proximity.  We can test for a number of typical indicators and scenarios.


\subsubsection{Planetary nature of the 92-day signal}

We first determine that the transit event is indeed occurring on TOI-1221 and not a nearby star blended in the \emph{TESS} aperture.
In NEB false positive scenarios, transiting stellar companions of comparable size to their host star are prone to producing V-shaped transits or grazing geometries, both of which are argued against by the flat bottom and clear demarcations of ingress and egress seen in the light curves.

We explicitly check the surrounding sky for potential NEB sources using small ground-based telescopes to clear all reasonably bright sources in the region.
In this context, clearing neighbours stars means that the unblended light curves of these stars show significant non-detections of transit-like signals at depths that would produce the observed flux drop in the blended {\it TESS} aperture.
The PEST observation clears stars within $\Delta$mag $\sim$ 7.5 in a $2\farcm5$ radius with the exception of the nearest neighbour ($13\farcs2$ away SWW) which was not cleanly resolved from the target.  
The subsequent LCOGT-1m observations simultaneously detected a transit  on-target and was able to resolve and rule out an NEB signal on the $13\farcs2$ neighbour. %
Table~\ref{Tab:PEST_NEBs} shows the NEB potential for reasonably bright stars within $2\farcm5$ 
ruled out by the PEST and LCOGT measurements.
With these small-telescope observations, we can conclude that sources from $2\farcm5$ out to $\sim$10\arcsec \ show no indication of NEB signals.

Turning to spectroscopy, we derive RVs from the CHIRON data of TOI-1221 using fitted line profiles of each spectrum, which were extracted using least-squares deconvolution (LSD) of the observed spectrum against synthetic templates \citep{Donati1997}. 
The CHIRON RVs (listed in Table \ref{Tab:chiron_RVs}) show no large velocity variation across the 2-year baseline and wide phase coverage.
With the relatively flat nature of the RVs gathered near pre- and post-transit quadratures, the $1\sigma$ and $3\sigma$ upper limits on the planet mass, given its period, are 1.1 M$_{\rm Jup}$ and 3.5 M$_{\rm Jup}$, respectively.  
While the RV measurements are not sensitive enough to rule out massive, outer planetary companions, they do rule out stellar-mass components in the system with orbital periods similar to the observed 92-day signals.
This is in line with the {\it Gaia} Renormalised Unit Weight Error (RUWE) measure of 1.11, which is consistent with expectations for a single star.  
%



As a final quantitative assessment of the planet's validity we used the \texttt{TRICERATOPS} (Tool for Rating
Interesting Candidate Exoplanets and Reliability Analysis of
Transits Originating from Proximate Stars) statistical validation package for {\it TESS} transits \citep{triceratops_code,triceratops_paper}.  
\texttt{TRICERATOPS} uses a Bayesian framework developed with {\it TESS} observations in mind.  It leverages the well-characterized stellar qualities in the {\it TESS} Input Catalog and specifically handles multiple blended stars in the large {\it TESS} apertures in a precise manner.
It calculates false positive probabilities by leveraging known information about background star rates and bound multiplicity priors.  It also applies provided contrast curves to refine its assessment.

\texttt{TRICERATOPS} calculates an
intrinsic false positive probability (FPP) which considers scenarios only involving absent or unresolved close neighbours,
as well as a nearby false positive probability (NFPP) which involves scenarios with nearby stars.
Based on comparisons to externally vetted TOIs, \texttt{TRICERATOPS} defines a planet candidate as validated if it achieves a FPP $<0.015$ and NFPP $<0.001$.
For our analysis at hand, we entered in the six detected {\it TESS} transits of TOI-1221.01 and the high-contrast imaging limits placed by SOAR and Zorro.  We calculate the values 
FPP $= 0.0014 \pm 0.0003$ 
and 
NFPP $= 0.0012 \pm 0.0004$.
The FPP statistic falls well below the \texttt{TRICERATOPS} validation threshold, and the NFPP straddles the threshold between "validated" and ``likely planet" designations.
It is worth noting that {\tt TRICERATOPS} does not have a way to include RV constraints in its assessment, nor does it know that we have cleared the closest neighbours of NEB potential.  Its findings, given these omissions, are likely very conservative.

In sections \ref{sSec:Transit_Fitting} and \ref{sSec:TTVs} we discuss the presence of transit timing variations (TTVs) in the observations.  This is additional qualitative evidence against the NEB hypothesis.  For the observed transits to be caused by an NEB, this NEB would have to exhibit TTVs and would therefore have to be a trinary (or more) system.

\begin{table}
\caption{CHIRON radial velocities of TOI-1221}
\label{Tab:chiron_RVs} \begin{tabular}{cccc}
  \hline
  ${\rm BJD_{TDB}}$ & RV & $\sigma_{\rm RV}$ & phase \\
   & (km s$^{-1}$) &  (km s$^{-1}$) & \\ 
  \hline
  
  2458775.86419  & 15.9562  &  0.0201 & 0.054 \\
  2459345.46777  & 16.0370  &  0.0221 & 0.267 \\
  2459491.84557  & 16.0263  &  0.0519 & 0.863 \\
\hline

 \end{tabular}
\end{table}




\subsubsection{Limits on additional system companions}

We further probe tighter separations than was possible with the small ground-based telescope observations.  The {\it Gaia} DR3 catalogue \citep{Gaia_collaboration_2016,Gaia_DR3_2022} 
reports no sources within 12\arcsec, and so clears the area down to its resolving power of $\lesssim$1$\arcsec$ with its high completeness limits at $\sim$17 Gaia magnitudes%
\footnote{\url{https://gea.esac.esa.int/archive/documentation/GEDR3/Catalogue_consolidation/chap_cu9val/sec_cu9val_introduction/ssec_cu9val_intro_completeness.html}}.
For the region within 1\arcsec, we look at the acquired high contrast speckle imaging.  
The initial SOAR/HRCam and subsequent Gemini-S/Zorro images clear the surrounding sky of sources (at a 5$\sigma$ level) down to $\Delta$mag of 5.2 at $0\farcs1$ separation.
This inner working angle corresponds to $\sim$13.7 au given the Gaia parallax distance of 137 pc. 
Out towards $3\arcsec$ the contrast improves to $\Delta {\rm mag} = 7$, as shown in the curves of Figure~\ref{fig:Speckle}.
These $\Delta$mag limits rule out the presence of anything more luminous than approximately an M4V dwarf at $0\farcs1$ or M6V dwarf out towards $1\arcsec$.

The CHIRON spectra are useful in this application as well.
Blended eclipsing binaries can produce spectral line profile variations as a function of orbital phase, but we find no evidence for additional stars or variable line profile shape in the spectra of TOI-1221.
Additionally, while the mostly flat RVs limit anything orbiting at $P\sim90$ days to planetary mass, a small, marginally significant RV slope exists ($0.13 \pm 0.053$ m s$^{-1}$ day$^{-1}$) that could indicate an additional massive object with a wider orbit.  
With only 3 data points, directly fitting any kind of model is impractical.
However, we can explore some limiting cases.
Let us restrict ourselves to simple circular edge-on orbits that could account for the observed slope.
The RVs exhibit this slope over their span of $\sim$2 years.  
A 2 $M_{\rm Jup}$ object on a 3 au orbit (5.2 year period) could produce the slope, but smaller orbits would have RV variations of too high frequency.
At the other end of the scale, a $0.24 M_\odot$ star ($\sim$M4V) at 30 au (165 year period) could also produce the slope, but at an angular separation of $0\farcs2$ it would be bright and separated enough to be detected in the Zorro image.
Any number of mass--period combinations in between these scenarios could produce the observed slope while remaining undetected by Zorro. 
Photometrically, none of the objects in this range are  bright enough to have a significant impact on our transit fits. Even an unresolved M4V star would only cause a minor flux dilution, affecting our radius value by $<$1\% which is smaller than our measured uncertainties.

Of course, realistic RV curves depend strongly on orbital inclination, phase, and eccentricity which are all unknown.  There is the additional possibility of the observed RV ``slope" not being a true gradual slope, but undersampled aliasing of a moderate-strength signal with period shorter than the 2-year coverage.  
Much uncertainty remains because we have few RV measurements with poor precision.



While the 91.7-day period signal passes all planetary validation checks, the data allow for the presence of another companion in the system.  The observations leave room for another body out to roughly 30 au and less than about 0.24 $M_\odot$.  Such a companion does not pose photometric dilution issues, but may have an impact in inducing TTVs.
With all these validation results taken together, we consider the practical false-positives ruled out.  From this point onward, we consider the transit events to be caused by a planet and fit the light curve following that assumption.  We will adopt the nomenclature of TOI-1221 b to refer to this planet moving forward.



\begin{figure}
\centering
\includegraphics[width=\linewidth,trim=100 70 80 90,clip]{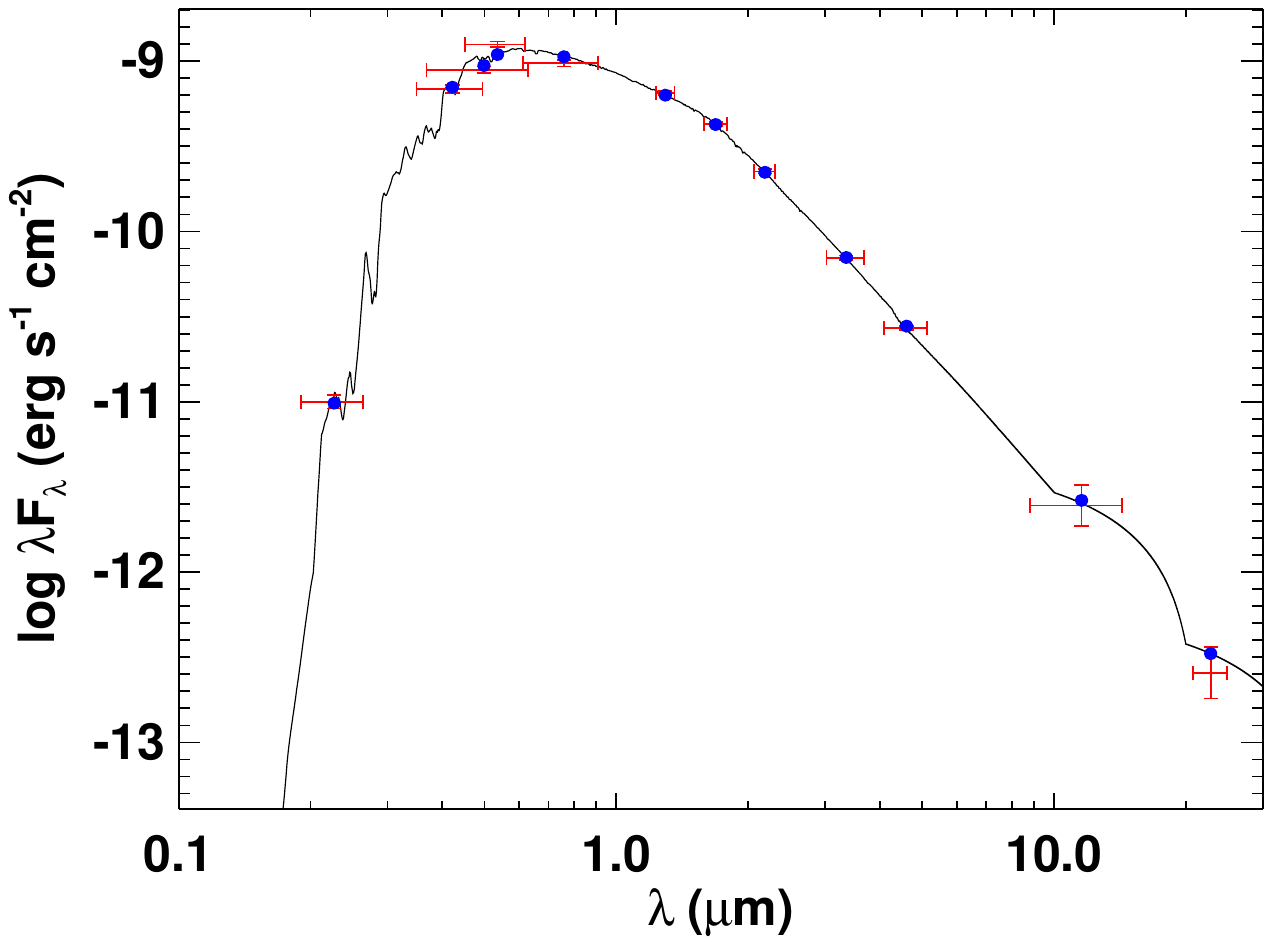}
\caption{
    Spectral energy distribution of TOI-1221. Red symbols represent the observed photometric measurements with the horizontal bars representing the effective width of the bandpass. Blue symbols are the model fluxes from the best-fit Kurucz atmosphere model (black).  \label{Fig:SED}}
\end{figure}

\begin{table}
\caption{Stellar characterization by CHIRON spectrosocpy and SED analysis}
\centering
\label{Tab:stellar_char}
\begin{tabular}{cccc}

\hline
Parameter &  CHIRON  &  SED   &  units \\
\hline

M$_\star$ & $0.93\pm0.22$$^{a}$ & $1.03 \pm 0.06$ & M$_\odot$ \\
R$_\star$ & $1.031\pm0.022$$^{a}$ & $0.993 \pm 0.037$ & R$_\odot$ \\
$\rho_\star$ & $1580\pm190$ & $1190 \pm 290$ & kg m$^{-3}$ \\
T$_{\rm eff}$ & $5592 \pm 50$ & $5700 \pm 100$ & K \\
${\rm[Fe/H]}$ & $0.06 \pm 0.08$ & $0.0 \pm 0.3$ & dex \\
$\log g$ & $4.38 \pm 0.10$ & -- & dex \\
$v\sin i$ & $2.1 \pm 0.4$ & -- & km ${\rm s}^{-1}$ \\
$F_{\rm bol}$ & -- & $1.597 \pm 0.037 \times 10^{-9}$ & erg s$^{-1}$ cm$^{-2}$ \\
$A_V$ & -- & $0.06 \pm 0.04$ & -- \\
$\log R'_{\rm HK}$ & -- & $-5.2 \pm 0.1$ & -- \\
$\tau_\star ^{b}$ & -- & $4.9^{+3.2}_{-2.5}$ & Gyr \\
$\tau_\star ^{c}$ & -- & $10 \pm 2$ & Gyr  \\

\hline

\end{tabular}
\raggedright
\newline\noindent 
\tablenotetext{a}{Derived using CHIRON values for $T_{\rm eff}$ and $\log g$ along with Gaia parallax distance (137 pc) and SED $F_{\rm bol}$.}
\tablenotetext{b}{Gyrochronology}
\tablenotetext{c}{Activity}

\end{table}


\subsection{Stellar Characterization}

We performed an analysis of the broadband spectral energy distribution (SED) of the star together with the {\it Gaia\/} EDR3 parallax \citep[with no systematic offset applied; see, e.g.,][]{StassunTorres:2021}, following the procedures described in \citet{Stassun:2016,Stassun:2017,Stassun:2018}.
We pulled the $B_T V_T$ magnitudes from {\it Tycho-2}, the $JHK_S$ magnitudes from {\it 2MASS}, the W1--W4 magnitudes from {\it WISE}, the $G_{\rm BP} G_{\rm RP}$ magnitudes from {\it Gaia}, and the NUV magnitude from {\it GALEX}. Together, the available photometry spans the full stellar SED over the wavelength range 0.2--22~$\mu$m.

We performed a fit using Kurucz stellar atmosphere models, with the free parameters being the effective temperature ($T_{\rm eff}$) and metallicity ([Fe/H]). The remaining free parameter is the extinction $A_V$, which we limited to the maximum line-of-sight value from the Galactic dust maps of \citet{Schlegel:1998}. 
The resulting fit (Figure~\ref{Fig:SED}) has a reduced $\chi^2$ of 1.5, with $A_V = 0.06 \pm 0.04$, $T_{\rm eff} = 5700 \pm 100$~K, and [Fe/H] = $0.0 \pm 0.3$. 
Integrating the (unreddened) model SED gives the bolometric flux at Earth, $F_{\rm bol} = 1.597 \pm 0.037 \times 10^{-9}$ erg~s$^{-1}$~cm$^{-2}$. 
Taking the $F_{\rm bol}$ and $T_{\rm eff}$ together with the {\it Gaia\/} parallax gives the stellar radius, $R_\star = 0.993 \pm 0.037$~R$_\odot$. In addition, we can estimate the stellar mass from the empirical relations of \citet{Torres:2010}, giving $M_\star = 1.03 \pm 0.06$~M$_\odot$.

Finally, we can use the star's NUV flux to estimate an age via empirical activity-age relations. The observed NUV excess implies a chromospheric activity of $\log R'_{\rm HK} = -5.2 \pm 0.1$ via the empirical relations of \citet{Findeisen:2011}. This estimated activity implies an age of $\tau_\star = 10 \pm 2$~Gyr via the empirical relations of \citet{Mamajek:2008}.
This age estimate is broadly consistent with that estimated from the star's rotation via gyrochronology relations \citep{Mamajek:2008}. From the spectroscopic $v\sin i_\star$ together with $R_\star$ we obtain a projected rotation period of $P_{\rm rot}/\sin i_\star = 25 \pm 5$~days, implying an age of $\tau_\star = 4.9^{+3.2}_{-2.5}$~Gyr.

We additionally verify the low activity of TOI-1221 with the 23,000 photometric data points from WASP-South.  In searching for rotational modulations using methods described in \citet{Maxted_etal_2011}, we constrain the variability to a 95\% confidence upper limit of 0.7 mmag.  This relative inactivity is nicely consistent with the old age estimate from the NUV.

As an independent determination of basic stellar parameters, we estimated spectroscopic stellar parameters by matching our CHIRON spectra against a library of observed spectra that have previously been classified by the Stellar Parameter Classification routines \citep[SPC;][]{Buchhave2012}. We interpolated to the best-matched values of effective temperature, surface gravity, and metallicity using a gradient-boosting regressor implemented in the \texttt{scikit-learn} python module, and found $T_{\rm eff} = 5592 \pm 50$\,K, $\log{g} = 4.38 \pm 0.10$, and ${\rm [Fe/H]} = 0.06 \pm 0.08$. The projected rotational velocity, $v\sin{i_\star}$, is derived following \citet{Gray2005} and \citet{Zhou2018}, by fitting broadening kernels to the instrumental, macroturbulent, and rotational line profiles. It is measured to be $2.1 \pm 0.4 {\rm ~km\,s}^{-1}$, but we note that this is smaller than the resolution element of the instrument and may be susceptible to systematic errors greater than the internal uncertainty quoted here.

We determine another mass and radius estimate using the CHIRON $T_{\rm eff}$ and $\log g$ values combined with the Gaia parallax distance and SED bolometric flux.
The derived stellar densities ($\rho_\star$) from the CHIRON and SED analysis are in agreement with one another.  They also agree with the {\it TESS} Input Catalog \citep[TIC;][]{TIC_2018} value used as a prior in or model fit, discussed in the next section.
The parameter values for both the CHIRON spectroscopy and SED analysis are summarized in Table~\ref{Tab:stellar_char}.

\startlongtable
\begin{deluxetable*}{llcc}
%
\tablehead{\colhead{~~~Parameter} & \colhead{Description (Units)} & \colhead{Value} & \colhead{Prior}}
\startdata
%
%
%
\multicolumn{4}{l}{Target Parameters:} \\
~~~~$P$\dotfill &  Period$^{a}$ (days)\dotfill    & $91.68278^{+0.00032}_{-0.00041}$     &  -- \\
~~~~$T_0$\dotfill &  Epoch reference$^{a}$ (BJD$_{\rm TDB}$)\dotfill         & $2458404.1791^{+0.0035}_{-0.0030}$  &  --  \\
~~~~$T_{14}$ \dotfill & Transit duration (hours) \dotfill & $8.12^{+1.01}_{-1.14}$    &   -- \\
~~~~$R_{p}/R_{\star}$\dotfill&  Planet radius in stellar radii\dotfill & $0.02679^{+0.00067}_{-0.00056}$    &  $\mathcal{U}[0.0,0.05]$  \\
~~~~$R_p$ \dotfill & Planet radius ($R_\oplus$)\dotfill & $2.91^{+0.13}_{-0.12}$    &   -- \\
~~~~$\delta$ \dotfill &Transit depth $(R_p/R_\star)^2$ \dotfill & $0.00072^{+0.00004}_{-0.00003}$    &   -- \\
~~~~$b$\dotfill &  Impact parameter\dotfill   & $0.25^{+0.22}_{-0.14}$          &  $\mathcal{U}[0.0,1.027]$  \\
~~~~$i$\dotfill            & Inclination (deg)\dotfill   & $89.75^{+0.16}_{-0.08}$   &  -- \\
~~~~$a/R_{\star}$\dotfill  & Semi-major axis in stellar radii \dotfill   & $83.7^{+4.7}_{-4.0}$      &  --  \\
~~~~$a$ \dotfill & Semi-major axis (au) \dotfill & $0.387^{+0.026}_{-0.023}$    &   -- \\
~~~~$e$\dotfill            & Eccentricity \dotfill   & $<0.21$                     &  -- \\
~~~~$\omega^\prime$\dotfill & Modified argument of pericentre$^{b}$ (deg) \dotfill  & $108^{+37}_{-43}$     &   -- \\
~~~~$\esinw$\dotfill   &  $e,\omega$ parameterization \dotfill    & $0.00^{+0.16}_{-0.26}$       & $\mathcal{U}[-1.0,1.0]$   \\
~~~~$\ecosw$\dotfill  &  $e,\omega$ parameterization\dotfill     & $0.01^{+0.40}_{-0.39}$       & $\mathcal{U}[-1.0,1.0]$   \\
~~~~$\rho_p$ \dotfill & Planet density$^{c}$ (g cm$^{-3}$) \dotfill & $2.01^{+0.71}_{-0.63}$   &   -- \\
~~~~$M_p$ \dotfill & Planet mass$^{c}$ ($M_\oplus$)\dotfill & $9.4^{+3.0}_{-3.7}$    &  -- \\
~~~~$\log g_p$ \dotfill & Surface Gravity$^{c}$ ($g_p$ given in cm s$^{-2}$) \dotfill & $3.15 ^{+0.19}_{-0.31}$   &   -- \\
~~~~$K$ \dotfill & RV semi-amplitude$^{c}$ \dotfill & $1.24^{+0.75}_{-0.91}$   &   -- \\
~~~~$S_p$ \dotfill & Insolation$^{d}$ ($S_\oplus$)\dotfill & $6.06^{+0.85}_{-0.77}$    &    -- \\
~~~~$T_{\rm eq}$ \dotfill & Equilibrium temperature$^{e}$ (K)\dotfill & $440\pm60$    &    -- \\
~~~~$\rho_{\star}$\dotfill&  Stellar density (kg m$^{-3}$)\dotfill & $1320^{+220}_{-190}$  &   $\mathcal{N}(1349, 296)$  \\
~~~~$T_{\#0}$\dotfill&  Epoch 0 transit midpoint (BJD$_{\rm TDB}$)\dotfill &$2458404.1844^{+0.0060}_{-0.0051}$ & $\mathcal{N}(2458404.1850, 0.02)$  \\
~~~~$T_{\#1}$\dotfill&  Epoch 1 transit midpoint (BJD$_{\rm TDB}$)\dotfill &$2458495.8474^{+0.0075}_{-0.0048}$ & $\mathcal{N}(2458495.8681, 0.02)$  \\
~~~~$T_{\#2}$\dotfill&  Epoch 2 transit midpoint (BJD$_{\rm TDB}$)\dotfill &$2458587.5401^{+0.0061}_{-0.0087}$ & $\mathcal{N}(2458587.5511, 0.02)$  \\
~~~~$T_{\#3}$\dotfill&  Epoch 3 transit midpoint (BJD$_{\rm TDB}$)\dotfill &$2458679.2361^{+0.0056}_{-0.0056}$ & $\mathcal{N}(2458679.2342, 0.02)$  \\
~~~~$T_{\#8}$\dotfill&  Epoch 8 transit midpoint (BJD$_{\rm TDB}$)\dotfill &$2459137.6494^{+0.0043}_{-0.0061}$ & $\mathcal{N}(2459137.6494, 0.02)$  \\
~~~~$T_{\#9}$\dotfill&  Epoch 9 transit midpoint (BJD$_{\rm TDB}$)\dotfill &$2459229.3342^{+0.0067}_{-0.0051}$ & $\mathcal{N}(2459229.3325, 0.02)$  \\
~~~~$T_{\#12}$\dotfill&  Epoch 12 transit midpoint (BJD$_{\rm TDB}$)\dotfill &$2459504.3456^{+0.0054}_{-0.0047}$ & $\mathcal{N}(2459504.3690, 0.02)$  \\
~~~~$T_{\#14}$\dotfill&  Epoch 14 transit midpoint (BJD$_{\rm TDB}$)\dotfill &$2459687.7466^{+0.0044}_{-0.0044}$ & $\mathcal{N}(2459687.7500, 0.02)$ \smallskip \\
\multicolumn{4}{l}{Telescope-specific Parameters:} \smallskip\\
~~~~$q_{\rm 1-TESS}$\dotfill&  Quadratic limb darkening coefficient$^{f}$\dotfill  & $0.26^{+0.02}_{-0.01}$          &  $\mathcal{N}(0.264,0.02)$     \\
~~~~$q_{\rm 2-TESS}$\dotfill&  Quadratic limb darkening coefficient$^{f}$\dotfill  & $0.34^{+0.01}_{-0.02}$          &  $\mathcal{N}(0.334,0.02)$     \\
~~~~$q_{\rm 1-LCOGT}$\dotfill&  Quadratic limb darkening coefficient$^{f}$\dotfill & $0.25^{+0.02}_{-0.03}$           &    $\mathcal{N}(0.250,0.02)$   \\
~~~~$q_{\rm 2-LCOGT}$\dotfill&  Quadratic limb darkening coefficient$^{f}$\dotfill & $0.30^{+0.02}_{-0.02}$            &    $\mathcal{N}(0.301,0.02)$   \\
~~~~$u_{\rm 1-TESS}$\dotfill& Quadratic limb darkening coefficient$^{f}$ \dotfill  & $0.35^{+0.02}_{-0.02}$    &  -- \\
~~~~$u_{\rm 2-TESS}$\dotfill& Quadratic limb darkening coefficient$^{f}$ \dotfill  & $0.16^{+0.02}_{-0.01}$   &  -- \\
~~~~$u_{\rm 1-LCOGT}$\dotfill& Quadratic limb darkening coefficient$^{f}$ \dotfill & $0.30^{+0.02}_{-0.02}$    &  -- \\
~~~~$u_{\rm 2-LCOGT}$\dotfill& Quadratic limb darkening coefficient$^{f}$ \dotfill & $0.20^{+0.01}_{-0.02}$    &  -- \\
~~~~$\mu_{\rm TESS}$ \dotfill & TESS relative flux offset \dotfill & $-0.00011 \pm 0.00001$   &  $\mathcal{N}(0.0,0.1)$ \\
~~~~$\mu_{\rm LCOGT}$ \dotfill & LCOGT relative flux offset \dotfill & $0.00003 \pm 0.00004$   &  $\mathcal{N}(0.0,0.1)$ \\
~~~~$\sigma_{GP-TESS}$ \dotfill & GP amplitude$^{g}$ (ppm) \dotfill & $<0.132$   &   $\mathcal{L}[10^{-6},10^2]$ \\
~~~~$\rho_{GP-TESS}$ \dotfill & GP time-scale of Matern part$^{g}$ (days) \dotfill & $<0.725$   &   $\mathcal{L}[10^{-6},10^2]$ \\
~~~~$\tau_{GP-TESS}$ \dotfill & GP time-scale of exponential part$^{g}$ (days) \dotfill & $<0.013$   &   $\mathcal{L}[10^{-6},10^2]$ \\
~~~~$\sigma_{GP-LCOGT}$ \dotfill & GP amplitude$^{g}$ (ppm) \dotfill & $<0.175$   &   $\mathcal{L}[10^{-6},10^2]$ \\
~~~~$\rho_{GP-LCOGT}$ \dotfill & GP time-scale of Matern part$^{g}$ (days) \dotfill & $<0.486$   &   $\mathcal{L}[10^{-6},10^2]$ \\
~~~~$\tau_{GP-LCOGT}$ \dotfill & GP time-scale of exponential part$^{g}$ (days) \dotfill & $<0.148$  &   $\mathcal{L}[10^{-6},10^2]$ \\ 
\enddata
\label{tab:full_results} 
\tablenotetext{}{%
%
Posted uncertainties (and limits) encompass the 68\% confidence interval.
Directly fitted parameters include a prior description.
Priors are of three varieties: Uniform over a given interval, $\mathcal{U}[a,b]$, normal with stated mean and standard deviation, $\mathcal{N}(\mu,\sigma)$, or log-uniform over a stated range, $\mathcal{L}[a,b]$.
If no prior is specified, the quantity was calculated from other parameters.
}
\tablenotetext{a}{%
$P$ and $T_0$ are determined as the best-fit linear ephemeris around which the individual transit timings vary.
}
\tablenotetext{b}{%
When eccentricity is low the light curve shape is insensitive to whether the transit occurs before or after pericentre passage. We define $\omega^\prime$ as the absolute angle between the line of sight and pericentre passage. This collapses $\omega$'s strongly bimodal distribution to a unimodal posterior and is easier to present in a written table.  $\omega$ has peaks at $198^\circ$ and $342^\circ$.
}
\tablenotetext{c}{%
Planetary mass estimate comes from from the model uncertainty of \citet[][]{Otegi_etal_2020}.  This value was used to calculate other mass-related quantities (e.g. $\rho_p$, $\log g_p$, and $K$).
}
%
\tablenotetext{d}{Insolation level assumes an orbital distance of $r=a$. }
\tablenotetext{e}{Assuming no albedo and perfect thermal redistribution.} 
\tablenotetext{f}{Definitions and discussion of the different parameterizations $\{u_1,u_2\}$ vs. $\{q_1,q_2\}$ can be found in \citet{Kipping_2013}.}
\tablenotetext{g}{The reported Gaussian process posterior limits all essentially reflect the input prior.}
\end{deluxetable*}





\begin{figure}
	\centering
	\includegraphics[width=0.47\textwidth]
    {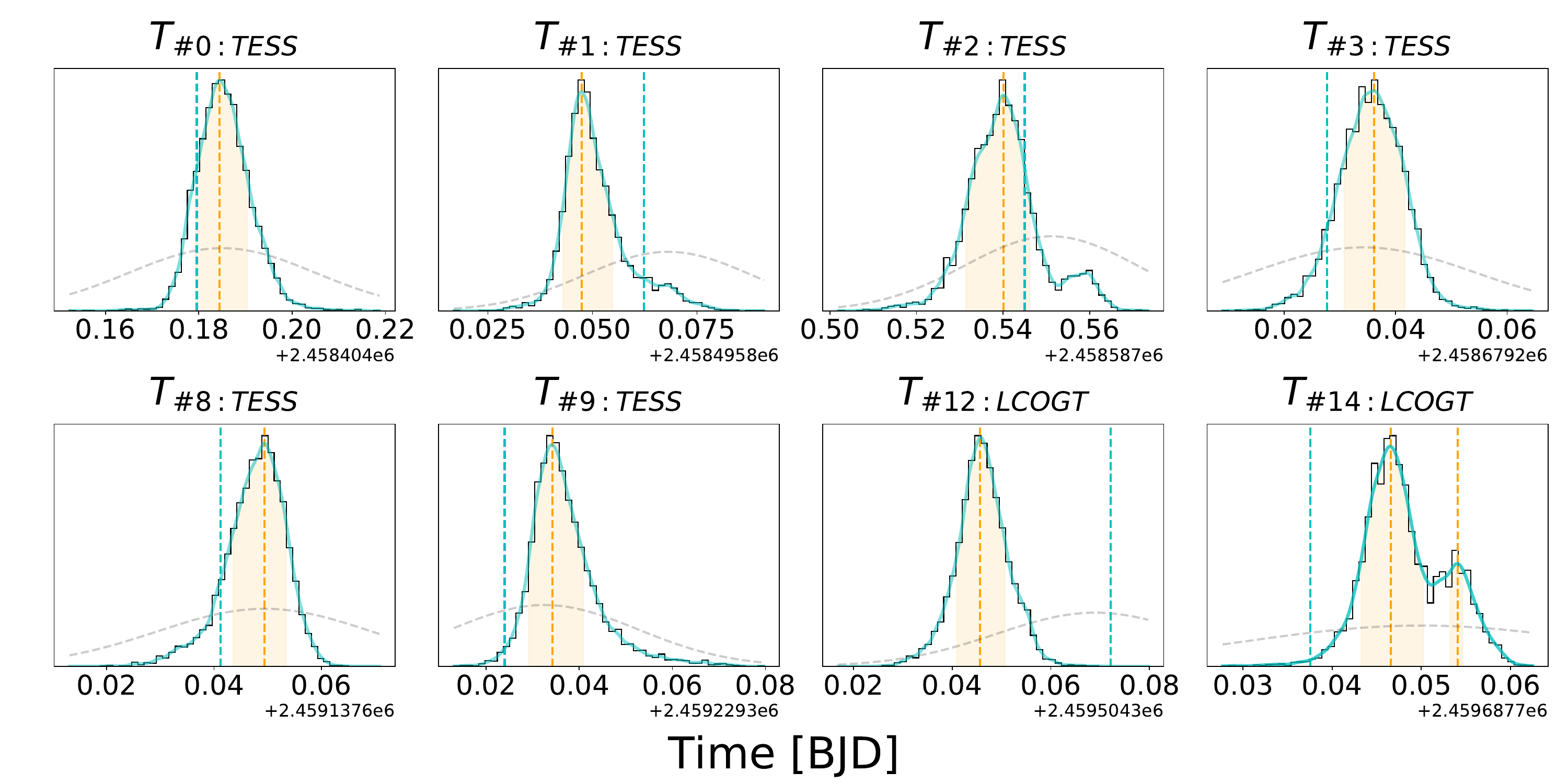}

    \caption{Individual transit midpoint posteriors for each transit in our data set.  In each plot, the orange region corresponds to the 68\% highest-posterior density (HPD) region, and the blue line shows where the best-fitting linear ephemeris would predict the midpoint to fall.  Grey dashed curves show the Gaussian prior placed on these midpoint timings.
        \label{fig:TTV_hists}}
\end{figure}

\subsection{Transit Fitting}\label{sSec:Transit_Fitting}

\subsubsection{Fitting with \juliet}\label{ssSec:juliet}

To extract planetary parameters from the light curve, we used the \juliet\ python package \citep{Espinoza_etal_2019}.
At its core, \juliet\ generates transit models using the BAsic Transit Model cAlculatioN  \citep[\batman;][]{Kreidberg_2015} code and includes a range of MCMC and nested sampling methods to fit photometric light curves and radial velocity data sets.  Among many other features, it can handle multiple planets, multiple instruments and data sets, and detrending via linear models or Gaussian processes.


The \batman\ transit modelling package requires the input parameters
$T_0, P, R_p/R_\star, a/R_\star, i, e, \omega$ as well as a limb darkening model and associated coefficients.  However, although it can uniquely generate a transit light curve with these parameters, many of them do not facilitate efficient exploration of physically relevant parameter space.  \juliet\ allows for a number of equivalent parameterizations that do a better job in this regard.  

One substitution we made was to forego directly fitting the semi-major axis ($a/R_\star$) in favour of fitting the stellar density ($\rho_\star$).
%
%
For a given period, the two are directly interchangeable using Kepler's 3rd Law, and there are independent constraints already set on the stellar density to use as a prior \citep{TIC_2018}.
%
We also re-parameterize the eccentricity ($e$) and argument of periastron ($\omega$) by instead fitting $\esinw$ and $\ecosw$ for which it is easier to efficiently sample the entire parameter space.
Finally, we employ a quadratic limb-darkening law and use the $q_1,q_2$ parameterization of \citet{Kipping_2013}, corresponding to the transformations in \citet{Espinoza_Jordan_2016} for $u_1,u_2$.  Again, this simply samples the allowable space with improved efficiency.


\subsubsection{TTVs}\label{ssSec:ttvs}


We found a few strange inconsistencies in our initial exploratory MCMC fits of the transit data.
One oddity was a curious bimodal distribution in the $T_0$ and $P$ posteriors.
Another was the tendency for walkers in longer MCMC runs to eventually transition from a model with small impact parameter and low eccentricity (which we will label $m_1$) to a model with much more extreme parameters ($m_2$). 
The extreme model $m_2$ differed primarily in a few key parameters.  The  impact parameter suggested a nearly grazing transit geometry ($b_2>0.9$), the eccentricity was drawn to very large values ($e_2>0.7$), and planetary radius came out about 18\% larger.
Despite the wild changes in these transit parameters, $m_1$ and $m_2$ produce very similar transit shapes. The only subtle difference between the two was a slight flattening of the ingress and egress slopes of $m_2$ compared to $m_1$.
Subsequent tests using nested sample routines found the same model shift even more easily, likely because nested sample techniques tend to be less prone to getting trapped in local likelihood maxima.

The bimodal $P$ and $T_0$ posteriors were are first clues that transit timing variations (TTVs) may be at play, and the specific differences in parameters between $m_1$ and $m_2$ lent credence to the idea.
As described in \citet[][Fig. 3]{Kipping2014}, the Photo-Timing Effect produces exactly these biases when existing TTVs are unaccounted for. 
If one tries to fold transit observations according to a single period, which is effectively how a fixed-period fit operates, then the presence of TTVs will wash out the ingress and egress signals to a degree when they do not line up properly.

In our case at hand, the artificially flattened ingress and egress slopes mimic a high-$b$ transit shape. A large impact parameter typically shortens the transit duration (by having a shorter chord across the stellar surface) unless another parameter can compensate.  
A shift of the semi-major axis ($a/R_\star$) could make up the difference, except that it is quite well-constrained by our prior knowledge of the stellar density ($\rho_\star$).  
The only remaining free parameter that could preserve the observed transit duration is a shift to high eccentricity (and accompanying appropriate argument of pericentre).
A high-$b$ geometry also implies the planet transits across the less bright limb of the star and possibly only grazes the stellar disk, requiring a larger planetary radius to affect the same observed flux drop.  
These are exactly the differences we see between $m_1$ and $m_2$ parameter sets.
We take this line of reasoning as additional qualitative evidence that TTVs are truly present in this data.

\begin{figure}
	\centering
	\includegraphics[width=0.47\textwidth]
    {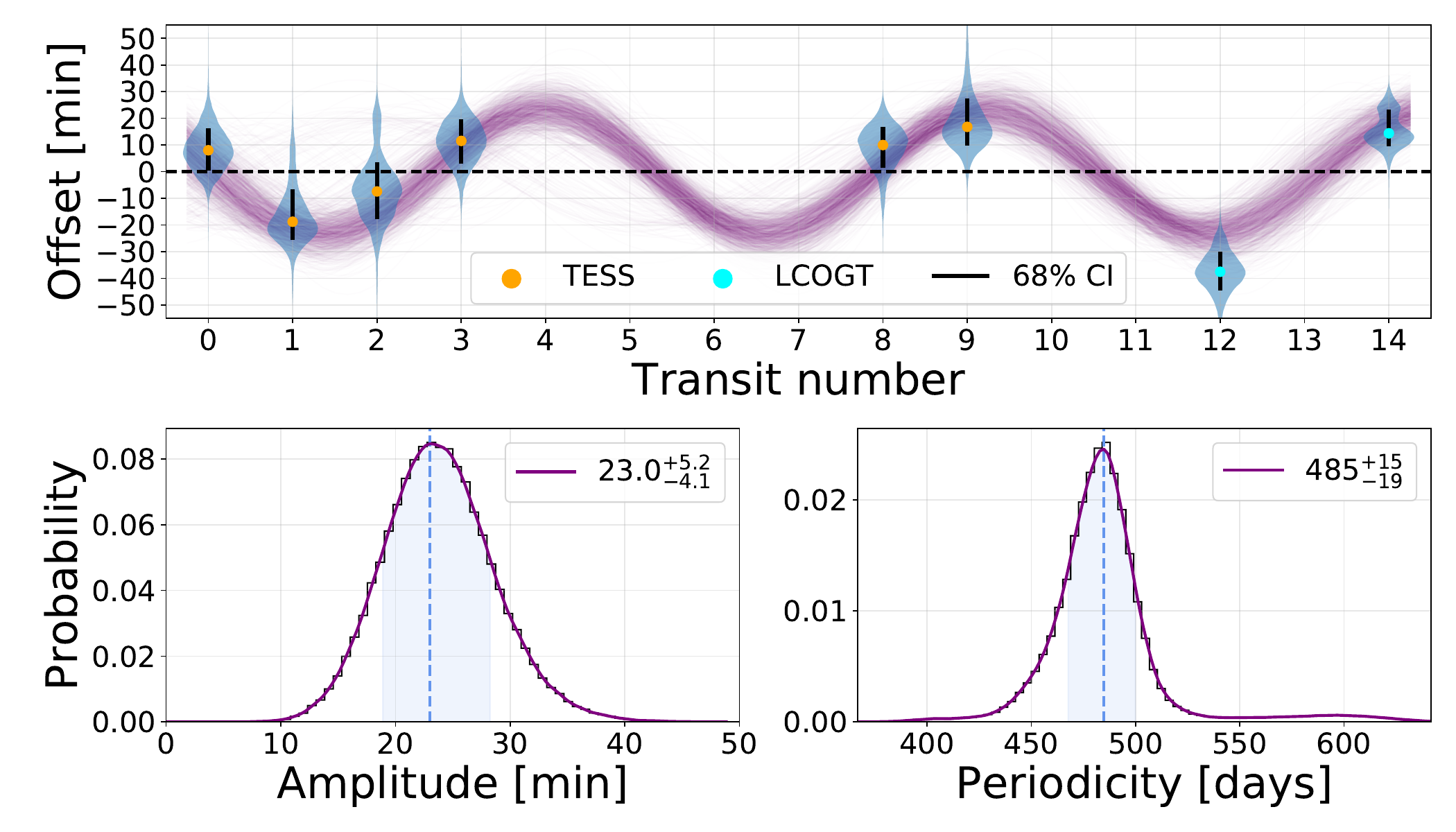}

    \caption{
    \emph{Top:} Midpoint timing distributions from Figure~\ref{fig:TTV_hists} plotted across time.  Width of the blue region indicates posterior probability distribution. The purple curves show sinusoid fits to 1000 random draws of these distributions.  The timings display a sinusoidal variation suggestive of systematic TTVs.
    \emph{Bottom:} Distributions of the amplitude and periodicity from fitting $10^5$ random draws.  A zero-amplitude line (i.e. fixed-period ephemeris) is disfavoured at $5.7\sigma$.
        \label{fig:TTV_O-C}}
\end{figure}

We decided that the most unbiased approach was to employ a transit fitting routine that made no assumptions of fixed periodicity.  
\juliet\ offers just such a functionality, allowing priors to be set on individual timings for each detected transit.  In this way, it is not simply the period that is fit, but the individual transit midpoints.
Other parameters are constant across all transits for a given sample model, but the normally fitted timing parameter $T_0$ is replaced by several $T_i$ parameters, one for each transit epoch.
General $P$ and $T_0$ posteriors are determined by \juliet\ , but these simply indicate the best linear ephemeris around which the TTVs oscillate.

We carried out the TTV fit on the combined set of the {\it TESS} and LCOGT observations shown in Figures~\ref{fig:TESS_data} \& \ref{fig:LCO_lc}
using \juliet's built-in nested sampling routines (\texttt{dynesty}, in this case).  
Extreme parameter models similar to $m_2$ are completely absent when we make this kind of timing-agnostic fit, and only results similar to the moderate $m_1$ model are favoured.
This leads us to believe that the $m_2$-like models are not physically motivated, but are the result of trying to force a fixed period model onto data with variable transit timing.  
The specific results of the TTV fit are addressed in Section~\ref{ssSec:fitresults}.


\subsubsection{Priors}\label{ssSec:priors}

Specific prior distributions for fitted parameters are listed in Table~\ref{tab:full_results}.  
Quantities without listed priors are not directly fitted, but are calculated from other values.
$P$ and $T_{\rm 0}$ do not receive specific priors in the typical sense.  Instead, each individual transit midpoint was given a wide
Gaussian prior (Table~\ref{tab:full_results} and Figure~\ref{fig:TTV_hists}) based on timing predictions from the {\it TESS} team's original fixed-period ephemeris. 
%
%
%
The planetary radius parameter $R_{\rm p}/R_\star$ receives a uniform prior over a range that comfortably encompasses the observed transit depth, including the null option of zero depth.  The uniform prior for $b$ covers the entire possible range given the expected planet size.  
Physically motivated limb darkening parameters $q_1$ and $q_2$ for both {\it TESS} and LCOGT were calculated using code by \citet{Espinoza_Jordan_2015} and the priors determined by the individual telescope response functions and uncertainties in stellar parameters.  
We used uninformative uniform priors over the whole range of the $\esinw$ and $\ecosw$ parameters, and a Gaussian prior for $\rho_\star$ which comes from the {\it TESS} Input Catalog \citep{TIC_2018}.
%
We also include a Gaussian process in the fit.  Amplitude and length-scale parameters were given very wide log-uniform priors spanning many orders of magnitude.


\begin{figure}
	\centering
	\includegraphics[width=0.48\textwidth,trim=30 10 100 10,clip] 
    {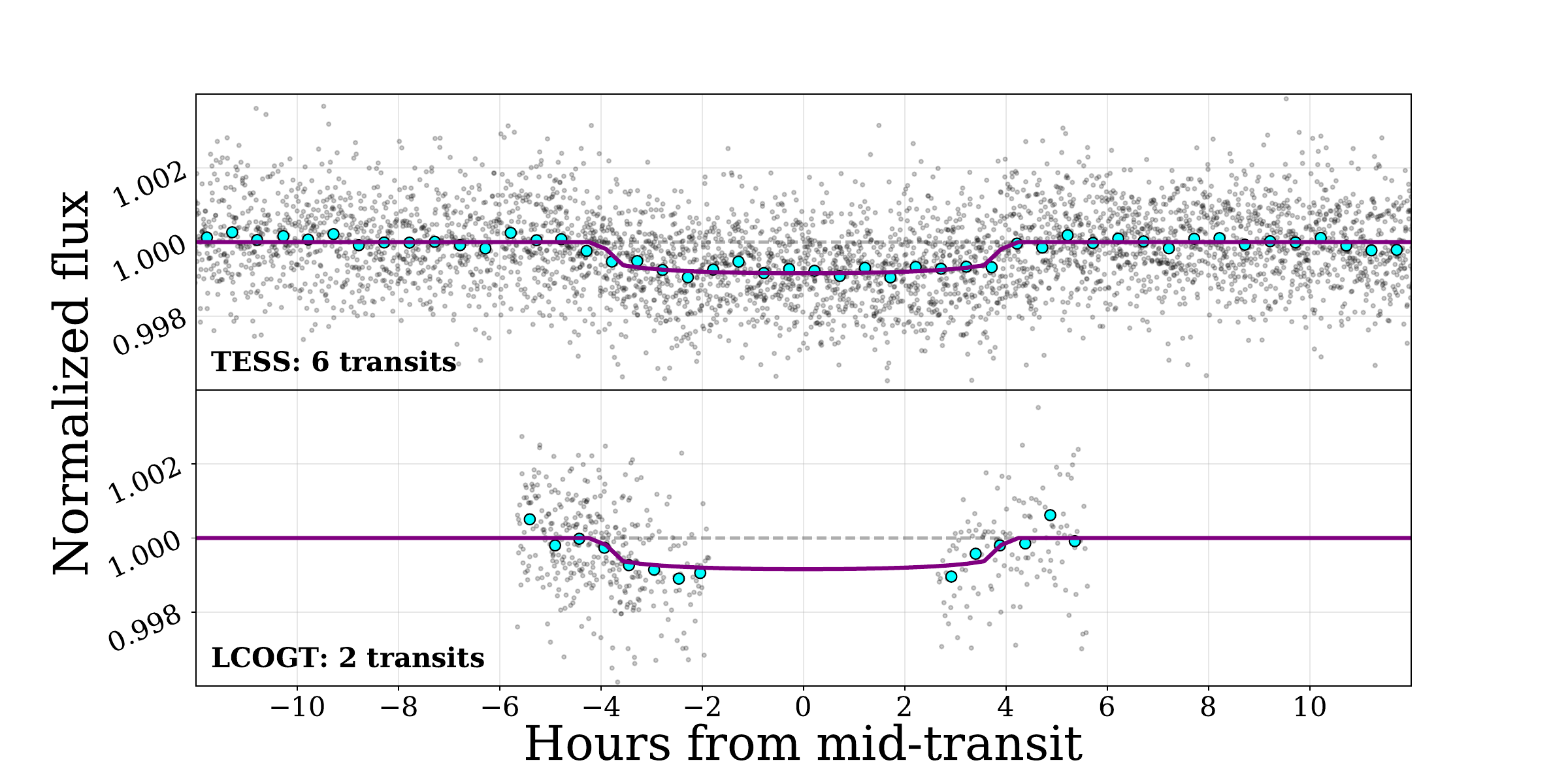}
    \caption{
    All {\it TESS} and LCOGT transits stacked and aligned to transit midpoints with highest likelihood model overlaid. Cyan points are data binned to 30 minutes.
    \label{fig:LC_final_model}}
\end{figure}



\subsubsection{Fit results} \label{ssSec:fitresults}


%
The fitted and derived parameter posterior distributions are summarized in Table \ref{tab:full_results}.
TOI-1221 b is approximately 
2.9
R$_{\earth}$ in size, or $\sim$75\%  Neptune's radius.
It orbits at 
$\sim$0.4
au and, receiving about 6 times the Earth's insolation and assuming no albedo and no tidal locking, would have an equilibrium temperature ($T_{\rm eq}$) of roughly 440~K.  
The timing posteriors of each of the 8 transits are also shown graphically in Figure~\ref{fig:TTV_hists}.  For most of them, the expected transit midpoints of the best-fitting fixed-period ephemeris (blue lines) fall in the wings of the distribution. 
Comparing the observed posteriors with the priors (grey curves) shows these features are data-driven and not strongly influenced by the prior. 
Plotting the observed-minus-calculated (O-C) timing differences as a function of transit epoch, we arrive at the values displayed in Figure~\ref{fig:TTV_O-C}.  
The timing offsets seem to preferentially follow an oscillatory pattern with amplitude 
$23.5^{+5.2}_{-4.1}$ minutes and 
periodicity of
$485^{+15}_{-19}$ days 
($5.29^{+0.17}_{-0.21}$ orbits).  
The amplitude distribution excludes a fixed-period ephemeris (i.e. zero amplitude) at $5.7\sigma$.
We note that the two final (LCOGT) transits appear to have large TTV offsets and might be thought to drive the sinusoidal pattern. 
Removing these points and running the same analysis on only the \emph{TESS} transits slightly lessens the fitted amplitude by  $\sim$4 minutes and the non-fixed-period significance to $3.6\sigma$.  
The fitted oscillatory TTV signal is strengthened by the two LCOGT points, but does not rely on them.

We included Gaussian process detrending parameters in our fits (trying a number of kernel models), but results consistently came back with amplitudes at least an order of magnitude smaller than the photometric scatter and transit depth.  
Our final fit includes an (approximate) Matern multiplied by exponential kernel, implemented via the {\tt celerite} package \citep{celerite}.
The highest log-likelihood model from the final fit is shown in Figure~\ref{fig:LC_final_model}.

\begin{figure*}[t]
	\centering
	\includegraphics[width=0.99\textwidth,trim=50 20 100 10,clip]
    {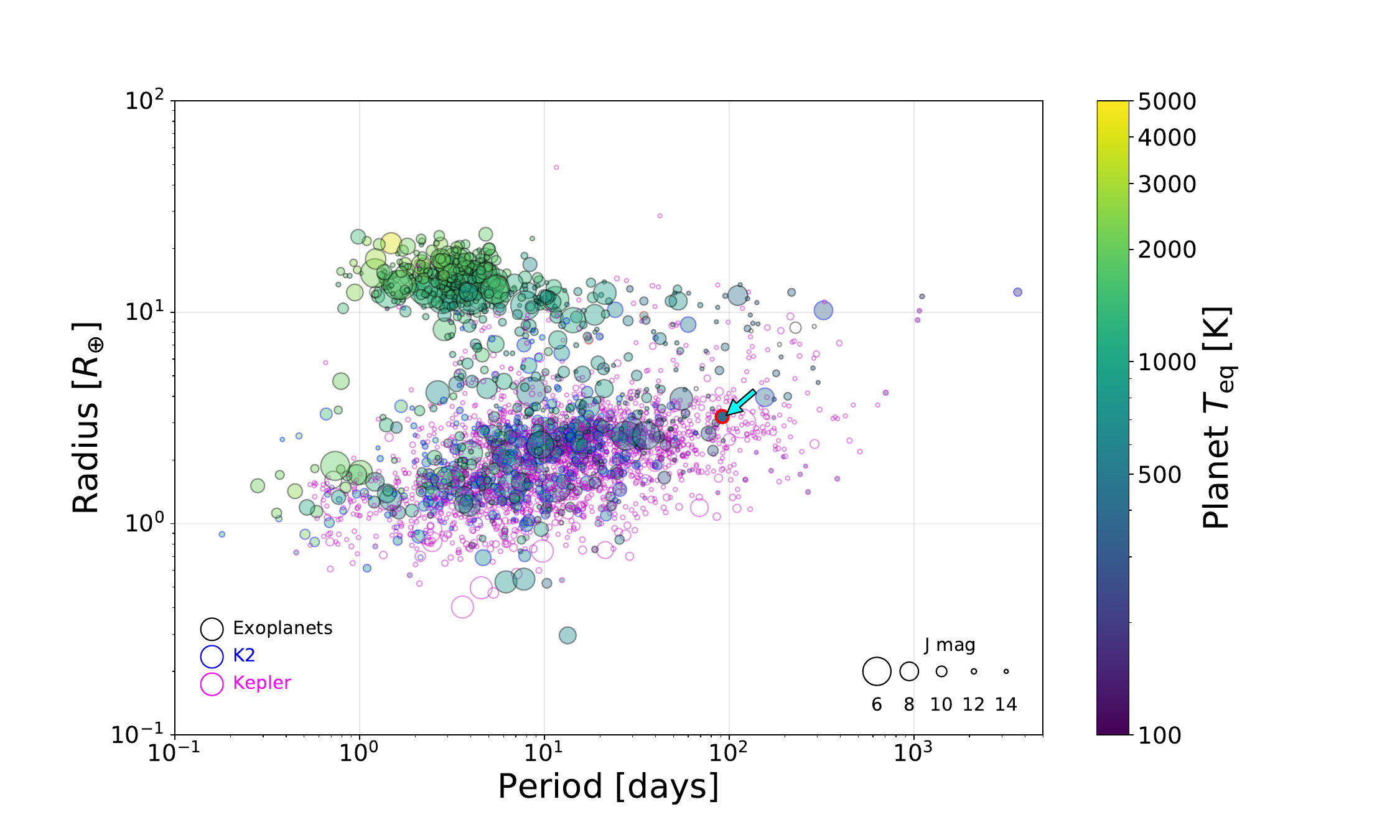}
    \caption{
    TOI-1221 b (highlighted with a red border and marked with the cyan arrow near $P=91$ days) placed in the context of other confirmed exoplanets with known radii. %
    Black marker outlines denote confirmed exoplanets with mass measurements.  
    Blue and magenta outlines are {\it K2} and {\it Kepler} statistically confirmed exoplanets without mass measurements, respectively.
    %
    \label{Fig:iExoView}}
\end{figure*}


\section{Discussion}\label{Sec:discussion}

\subsection{Planet properties}

In one sense, given its sub-Neptune size of 2.9 $R_\oplus$, TOI-1221 b is a very typical exoplanet, as super-Earths and sub-Neptunes are the most populous exoplanet size classes 
 \citep{NEXA_Planetary_Systems_Table}.
However, it has certain other traits that place it in the outskirts.
Among exoplanets that have well-constrained radii, TOI-1221 b is currently at the 97$^{\rm th}$-percentile for orbital period, and the 10$^{\rm th}$-percentile for $T_{\rm eq}$, placing it at the edges of this parameter space \citep[Figure~\ref{Fig:iExoView};][]{NEXA_Planetary_Systems_Table}.
This combination of a rare (or at least currently under-sampled) temperature regime on a common type of planet makes TOI-1221 b a valuable test-bed for understanding cooler planets.  

In the course of our analysis, we experienced first-hand the importance that relatively minor TTVs can have on the outcome of a traditional transit fit.  
The Photo-Timing Effect \citep{Kipping2014} caused dramatically different fit results, even when our TTVs are of a 20-minute scale compared to an 8-hour transit.
Allowing for some flexibility in transit timings as a test is a good practice whenever TTVs may be present.


\subsection{Transit Timing Variations}
\label{sSec:TTVs}

When plotted against transit epoch, the TTVs (Figures~\ref{fig:TTV_O-C} and \ref{fig:TTV_hists}) seem to adhere to a sinusoidal pattern.
An oscillatory pattern can be indicative of gravitational perturbations of an additional unseen planet, with the strongest effects near a first-order orbital resonance \citep{Lithwick_etal_2012}.  
Our analysis in Section~\ref{sec:RejectingFPs} of the mild RV slope and high-contrast imaging places some limits on this possible companion.  The RV slope could not be produced by an object smaller than  2 $M_{\rm Jup}$ on an orbit smaller than 3 au, and the Zorro imaging would detect an object more massive than an 0.24 $M_\odot$ orbiting at 30 au (which could also cause the RV slope).  However, objects of lower mass could potentially hide below our RV and imaging thresholds.
%


To explore what kind of unseen additional companion could cause the observed TTVs, we made use of the python module {\tt TTVFast} \citep{ttvfast}. 
{\tt TTVFast} accepts a list of bodies with specified masses and orbital elements, then calculates the expected transit times of each given the mutual gravitational interactions of the bodies involved.

A few findings became clear as we explored the effects of various perturbing objects on the transit timings of TOI-1221 b.
Firstly, TTVs generally have multiple components to them.  The major component is a long-period variation with large amplitude that typically oscillates over tens to hundreds or even thousands of transit epochs.  
The periodicity of this component is often referred to as the ``super-period" of the system's TTVs.  
A secondary component caused by more frequent synodic interactions is referred to as ``chopping" and has lower amplitude and higher frequency than the super-period signal.  
Our observed TTVs are more likely to be this short-period chopping signal, rather than the super-period of the system.

Secondly, we find that many different mass--orbit configurations of a secondary planet can faithfully reproduce the observed TTVs.  
These configurations can have dramatically different planet masses and orbital periods from one another, spanning all through the parameter space that remains ``invisible" to our data sets.
Unfortunately we also find that very small changes in the perturber's parameter space can produce wild changes in the resulting TTVs.  
The jagged complexity of the many-dimensional likelihood landscape means no single model gives uniquely satisfying results, no matter how closely it matches the observations. 

The {\tt TTVFast} exploration shows that it is certainly possible for another object in the system to create the TTVs we observe (particularly as a chopping signal), though they do not help very much in determining limits on the object's mass or orbit.  
There are many cases where TTVs can be used to estimate planet masses 
(e.g. \citet{Grimm_etal_2018} for TRAPPIST-1
or recently \citet{Greklek-McKeon_etal_2023} for Kepler-289), 
and chopping signals can even be used to break mass--eccentricity degeneracies \citep{chopping}.  
However, such cases generally require coverage of many more epochs than we have for TOI-1221 b (often dozens), and have the period or even TTV measurements of the other perturbing planets.  
The low SNR \emph{TESS} data inflating the timing uncertainties, the long 92-day period, and the lack of other detected transiting objects work against TOI-1221 b in this case.
Additional transits in upcoming \emph{TESS} sectors will help to extend the epoch baseline and may potentially show the beginnings of a super-period trend.

%
%

We also looked into transit duration variations (TDVs) that may accompany TTVs.  {\tt TTVFast} predicts small TDVs for most mass--orbit configurations, on the order of 1-10 minutes.  Given the low SNR of single transits, individual fits of our data produce duration uncertainties on the order of $\sim$30 minutes.  We do not achieve the timing precision necessary to draw conclusions about TDVs.

For the time being, we can only suggest that future searches of light curves or RV data sets on this target keep an eye out for additional signals, particularly near first-order interior and exterior resonances
(e.g. periods of 2:1--45.84, 3:2--61.12, 4:3--68.76, 5:4--73.35,
 4:5--114.60, 3:4--122.24, 2:3--137.52, and 1:2--183.36 days).


\subsection{Future Observations}
\label{sSec:future_obs}

A full confirmation with an RV mass measurement is the obvious next step for TOI-1221 b.
With our fitted planetary radius, we can calculate the expected RV semi-amplitude signal ($K$) given some assumption about the planet's bulk density.  
With the mass estimate from \citet{Otegi_etal_2020} (displayed in Table~\ref{tab:full_results}), we expect $K\sim1.2$ m s$^{-1}$, a challenging detection.
Besides providing a mass, a precision RV campaign would also allow for a refinement of the scale height estimate and associated atmospheric signal strength. It could potentially also detect a signal from a perturbing body causing the observed TTVs.

If we assume a mean molecular weight similar to Neptune
($\mu = 2.6$ u), 
TOI-1221 b's atmospheric signal strength (Eq. 6.154 of \citet{Perryman2018}) is predicted to be around 50 ppm during transit.
This small value is an expected challenge for cooler planets, but is not outside the realm of possibility with the James Webb Space Telescope (JWST), especially as the target is relatively bright ($H,J,K\sim9$). 
Using the transmission spectroscopy metric (TSM) of \citet[][Eq. 1]{Kempton_etal_2018} we find a value of
$19.4$.
The TSM represents the expected signal-to-noise ratio for a 10-hour JWST/NIRISS spectroscopic observation.
This places TOI-1221 b slightly above the middle range of TSM values for other TOIs with reliable periods greater than 50 days (roughly $15 \pm 9$).

TOI-1221 b fits the {\it TESS} Mission Level One Science Requirement of measuring the mass of 50 planets with radii smaller than 4 R$_{\earth}$.
Though the goal was officially  reached in Oct 2022, this does not diminish the value of targets like TOI-1221 b.  These are the kinds of planets {\it TESS} was designed to find.
Longer-period and cooler planets generally require more time and effort to uncover and verify than their short-period counterparts, yet they allow us to probe very different atmospheric regimes.

While a plethora of information exists for more easily accessible short-period planets, only recently are we beginning to produce detailed examples of their longer-period brethren.  We are pushing those boundaries towards the regime where we can make meaningful comparisons with our own slow and cold solar system bodies.
Given TOI-1221 b's reasonable TSM value, it provides a good experimental case for the effects of low insolation on a ubiquitous planet class.

\section{Summary \& Conclusion}\label{Sec:summary}


In the course of this paper, we have validated the exoplanet TOI-1221 b.
In doing so we observed and collated a wide range of data sets. 
Ground-based time series photometry observations from small telescopes were able to determine that the transit signal was not caused by an eclipsing binary in any of the nearby stars.  
The possibility of an on-sky blend due to a background or foreground object was severely restricted by high contrast imaging by SOAR and Gemini-S/Zorro speckle images.
Any very low-mass M-dwarfs that could remain undetected in our data are not sufficiently luminous to cause dilution issues with the transit fit.
We further determine a low false positive probability with {\tt TRICERATOPS}.

We characterize the stellar properties both with spectra from CHIRON and an SED analysis based on archival photometric measurements.  The stellar values from these two methods are in line with one another and also with existing assessments \citep[e.g. the TIC;][]{TIC_2018}.

Straightforward transit fitting methods of the {\it TESS} light curves and two LCOGT transit observations were initially confounded by what appeared to be variations in the transit timings.  
Fitting routines that did not enforce a fixed-period on the model produced clearer results and revealed an oscillatory offset pattern in observed transit midpoints.
%
This is potentially indicative of perturbations by an additional unseen planet in the system, likely near a first-order period resonance.
Using {\tt TTVFast} we explored the TTVs induced upon TOI-1221 b by a wide range of additional planet mass--orbit configurations.  We determine that the observed TTVs are likely to be a high-frequency chopping signal, rather than the high amplitude but low frequency super-period.  
Unfortunately, a large number of wildly different configurations can all reproduce our observations with high accuracy.
Without transit detections of the perturber,  more TTVs of planet b over a much longer baseline are needed to make informative assessments of a perturber's mass and orbit.

Based on all of the work described above, we now consider TOI-1221 b to be a validated planet.  All of the likely false-positive are ruled out, and the transit parameters converge to reasonable values.  
Moving forward, a precision RV mass measurement would solidify TOI-1221 b's planetary status, and with luck may also reveal a signal of the unseen perturber acting in the system.

The origin of the super-Earth/sub-Neptune valley remains an important question.  Properly testing the various hypotheses likely requires a more thorough understanding of the effects of insolation on planetary atmospheres.  Building such a robust understanding will require analysis of the hard-to-find cool planets in addition to the more easily discovered and plentiful hot ones.
Of TOIs that have strong period constraints, TOI-1221 b is the 19$^{\rm th}$-longest period {\it TESS} planet detected to date (out of 6133 TOIs) and is pushing the boundaries of our populated parameter space.

%


\section*{Acknowledgements}

%
C.M. and D.L. acknowledge funding from 
the Trottier Family Foundation in their support of Trottier Institute for Research on Exoplanets (iREx).  
They also acknowledge individual funding from the Natural Sciences and Engineering Research Council (NSERC) of Canada.
D. D. acknowledges support from the TESS Guest Investigator Program grants 80NSSC21K0108 and 80NSSC22K0185.
S.Q. acknowledges support from the {\it TESS} GI Program under award 80NSSC21K1056.
M.H.K. thanks Allan R. Schmitt for making his light curve examining software {\tt LcTools} freely available.
The work of H.P.O. has been carried out within the framework of the National Centre of Competence in Research PlanetS supported by the Swiss National Science Foundation under grants 51NF40\_182901 and 51NF40\_205606 for which we acknowledge the financial support of the SNSF.
This paper made use of data collected by the {\it TESS} mission and are publicly available from the Mikulski Archive for Space Telescopes (MAST) operated by the Space Telescope Science Institute (STScI). Funding for the {\it TESS} mission is provided by NASA’s Science Mission Directorate.
We acknowledge the use of public {\it TESS} data from pipelines at the {\it TESS} Science Office and at the {\it TESS} Science Processing Operations Center.
Resources supporting this work were provided by the NASA High-End Computing (HEC) Program through the NASA Advanced Supercomputing (NAS) Division at Ames Research Center for the production of the SPOC data products.
This work has set the stage for a follow-up mass measurement of TOI-1221 b, a planet that matches the {\it TESS} Mission Level One Science Requirement of determining the masses of fifty (50) planets with radii less than 4 R$_{\earth}$.
This research has made use of the Exoplanet Follow-up Observation Program website, which is operated by the California Institute of Technology, under contract with the National Aeronautics and Space Administration under the Exoplanet Exploration Program.
%
%
This work makes use of observations from the LCOGT network. Part of the LCOGT telescope time was granted by NOIRLab through the Mid-Scale Innovations Program (MSIP). MSIP is funded by NSF.
This paper was based in part on observations obtained at the Southern Astrophysical Research (SOAR) telescope, which is a joint project of the Minist\'{e}rio da Ci\^{e}ncia, Tecnologia e Inova\c{c}\~{o}es (MCTI/LNA) do Brasil, the US National Science Foundation’s NOIRLab, the University of North Carolina at Chapel Hill (UNC), and Michigan State University (MSU).
Observations in the paper (Program ID: GN/S-2021A-LP-105) made use of the High-Resolution Imaging instrument Zorro. Zorro was funded by the NASA Exoplanet Exploration Program and built at the NASA Ames Research Center by Steve B. Howell, Nic Scott, Elliott P. Horch, and Emmett Quigley. Zorro was mounted on the Gemini South telescope of the international Gemini Observatory, a program of NSF’s NOIRLab, which is managed by the Association of Universities for Research in Astronomy  (AURA) under a cooperative agreement with the National Science Foundation on behalf of the Gemini Observatory partnership: the National Science Foundation (United States), National Research Council (Canada), Agencia Nacional de Investigaci\'{o}n y Desarrollo (Chile), Ministerio de Ciencia, Tecnolog\'{i}a e Innovaci\'{o}n (Argentina), Minist\'{e}rio da Ci\^{e}ncia, Tecnologia, Inova\c{c}\~{o}es e Comunica\c{c}\~{o}es (Brazil), and Korea Astronomy and Space Science Institute (Republic of Korea).
This research has made use of the NASA Exoplanet Archive, which is operated by the California Institute of Technology, under contract with the National Aeronautics and Space Administration under the Exoplanet Exploration Program.
This work has made use of data from the European Space Agency  {\it Gaia} (\url{https://www.cosmos.esa.int/gaia}), processed Data Processing and Analysis Consortium (DPAC, \url{https://www.cosmos.esa.int/web/gaia/dpac/consortium}). Funding for the DPAC has been provided by national institutions, in particular the institutions participating in the {\it Gaia} Multilateral Agreement.
%


\section*{Data Availability}

The {\it TESS} photometric light curves are freely available for download from the MAST archive (doi:\dataset[10.17909/t9-nmc8-f686]{http://dx.doi.org/10.17909/t9-nmc8-f686}).

The observations carried out by the TFOP group (PEST/ST-8XME, LCOGT/Sinistro, SMARTS/CHIRON, and Gemini-S/Zorro) have been uploaded to the group's online repository (doi:\dataset[10.26134/ExoFOP3]{https://exofop.ipac.caltech.edu/tess/target.php?id=349095149}). These data become publicly available after a 12-month proprietary period starting from the date of submission.


The {\it Gaia} DR3 data in the vicinity of TOI-1221 in this research is publicly available by searching the target coordinates on the online {\it Gaia} archive
(doi:\dataset[10.5270/esa-qa4lep3]{https://doi.org/10.5270/esa-qa4lep3}).

We pulled the $B_T V_T$ magnitudes from {\it Tycho-2}, the $JHK_S$ magnitudes from {\it 2MASS}, the W1--W4 magnitudes from {\it WISE}, the $G_{\rm BP} G_{\rm RP}$ magnitudes from {\it Gaia}, and the NUV magnitude from {\it GALEX}.

The WASP data are not available on a public archive, but can be provided upon request by contacting Coel Hellier
(c.hellier@keele.ac.uk).

\software{
{\tt Tapir} \citep{Jensen_2013}, 
{\tt BANZAI} \citep{banzai,McCully_etal_2018},
{\tt AstroImageJ} \citep{Collins_2017},
{\tt LcTools} \citep{lctools},
{\tt TRICERATOPS} \citep{triceratops_code,triceratops_paper}, 
{\tt scikit-learn} \citep{scikit-learn},
{\tt Juliet} \citep{Espinoza_etal_2019}, 
{\tt batman} \citep{Kreidberg_2015}, 
{\tt dynesty} \citep{dynesty_software,dynesty_paper},
{\tt celerite} \citep{celerite},
{\tt C-Munipack} \citep{C_Munipack}.
}



\bibliography{main.bib} 








\end{document}